\documentclass[11pt]{article}
\usepackage{amsfonts}
 \usepackage{amsmath}

\usepackage{mathrsfs}
\usepackage{fancyhdr}
\usepackage{cite}
\usepackage{color}
\usepackage{hyperref}
\usepackage[numbers,sort&compress]{natbib}
\newcommand{\bblu}{\begin{color}{blue}}
\newcommand{\bred}{\begin{color}{red}}
\newcommand{\ecl}{\end{color}}

\begin{document}
\renewcommand{\theequation}{\thesection.\arabic{equation}}

\begin{center}
\LARGE\bf Integrable symplectic maps associated with discrete Korteweg-de Vries-type equations
\end{center}

\begin{center}
 Xiaoxue Xu$^{\rm 1}$, Mengmeng Jiang$^{\rm 1}$, Frank W Nijhoff$^{\rm 2}$
\end{center}

\begin{center}
${}^{\rm 1}$ School of Mathematics and Statistics, Zhengzhou University, Zhengzhou 450001, PR China\\
${}^{\rm 2}$ Department of Applied Mathematics, University of Leeds, Leeds LS2 9JT, UK\\
\end{center}

\begin{abstract}
In this paper we present novel integrable symplectic maps, associated with ordinary difference equations, and show how they
determine, in a remarkably diverse manner, the integrability, including Lax pairs and the explicit solutions,
for integrable partial difference equations which are the discrete counterparts of integrable partial differential equations
of Korteweg-de Vries-type (KdV-type).
As a consequence it is demonstrated that several distinct Hamiltonian systems lead to one and the same difference equation
by means of the Liouville integrability framework. Thus, these
integrable symplectic maps may provide an efficient
tool for characterizing, and determining the integrability of, partial difference equations.
\end{abstract}


\section{Introduction }

Integrable symplectic maps \cite{Quispel,Bruschi,Suris,Veselov} comprise some of the remarkable outcomes from the theory of discrete integrable systems:
such maps allow the construction of special solutions for the corresponding  partial difference equations by means of algebro-geometric methods \cite{Belokolos,Matveev,Gesztesy,Cao7,Geng0}.
 Many of the infinite-dimensional discrete integrable models that are supported by (in the sense that they can be reduced to) integrable symplectic maps
have interesting properties: the existence of Lax pairs, B\"acklund transformations, symmetries and conservation laws, Hamiltonian structures, the construction of integrable algorithms, (elliptic) soliton solutions, finite genus solutions \cite[see][and references therein]{Cao,Hietarinta,Nijhoff,Zhu,Nijhoff1,Chang,Xenitidis,Zhang,Zhang0,Zhang1,Geng,Lou,Hu,Ma}.

By definition, in the symplectic space $\mathcal{N}=(\mathbb{R}^{2N},\mathrm{d}p\wedge \mathrm{d}q)$ with associated coordinates $(p,q)=(p_{1},\ldots,p_{N},q_{1},\ldots,q_{N})^{T}$, where $N$ is a positive integer, a mapping $S$ that sends $(p,q)$ to $(\tilde{p},\tilde{q})$ is called an integrable symplectic map, iff the induced map $S^{*}$ on the space of differential forms on $\mathcal{N}$ satisfies $S^{*}(\mathrm{d}p\wedge \mathrm{d}q)=\mathrm{d}p\wedge \mathrm{d}q$ and $S^{*}F_{j}=F_{j},1\leq j\leq N$, where $F_{1},\ldots,F_{N}$  are smooth, functionally independent, and pairwise in involution (with respect to the Poisson bracket associated with symplectic form), functions in a dense
open subset of $\mathbb{R}^{2N}$. We will construct in this paper some novel integrable symplectic maps, which turn out to constitute a powerful tool to study the natural discrete analogues of KdV-type equations. The latter are
members of the celebrated Adler-Bobenko-Suris (ABS) list \cite{Adler} of integrable partial difference equations on the quadrilateral lattice.

In \cite{Papageorgiou,Capel1}, some of the integrability characteristics of multidimensional mappings arising by periodic reductions from the partial difference analogues of the KdV equation were investigated.
These include Lax pairs, classical $r$-matrix structures and Liouville integrability, but the explicit solutions for the map remained to be established.
In \cite{Enolskii}, by means of the finite-gap technique, the rational maps generated from periodic initial problems of the lattice KdV equation, were parameterized in terms of
Kleinian functions, and the closed-form modified Hamiltonians for one- and two-degree symplectic mappings arising from the lattice KdV and modified KdV equations were
constructed, combined with the application of the  method of separation of variables \cite{Dubrovin,Nijhoff2,Alsallami}. The latter gave rise to a discrete analogue
of the Kowalewski-Dubrovin equations, describing the dynamics in terms of the separation variables. The quantization of the latter systems have been investigated as well
in connection with integrable quantum field systems \cite{Smirnov,Babelon0,Field}. Symplectic mappings related to higher-order counterparts of the KdV type, e.g., the Boussinesq type, have been studied as well \cite{Alsallami1}.

The motivation for this present paper originates from \cite{Cao}, in which finite genus solutions for the lattice potential KdV (lpKdV) equation
\begin{equation}\label{eq:1.1}
(\bar{\tilde{u}}-u)(\tilde{u}-\bar{u})=\beta_{2}-\beta_{1},
\end{equation}
are obtained through integrable symplectic maps and where the lattice KdV (lKdV) equation
is solved as well. Here we use the usual notation, $\tilde{h}(m,n)=h(m+1,n),
\bar{h}(m,n)=h(m,n+1)$, for any function $h(m,n)$. A discrete spectral problem associated with \eqref{eq:1.1} can be constructed from the property of multi-dimensional consistency of \eqref{eq:1.1}, cf. \cite {Hietarinta}, namely
\begin{equation}\label{eq:1.2}
\tilde{\chi}=(\lambda-\beta)^{-1/2}D^{(\beta)}(\lambda;a,b)\chi,\ \ D^{(\beta)}(\lambda;a,b)=\begin{pmatrix}
 a & -\lambda+\beta+ab \\
1 &b
\end{pmatrix}\chi,
\end{equation}
where $\chi$ is a 2-component vector function, $\lambda$ is a spectral parameter, $\beta$ is the parameter of
the lattice and $a,b$ are potentials (i.e., functions of the independent variables), which is different from the ones in \cite{Cao,Capel}. Furthermore, from the Darboux/B\"acklund approach, it is found that \eqref{eq:1.2} also allows a compatible spectral problem \cite{Rogers,Levi,Li}
\begin{equation}\label{eq:1.3}
\partial_{x}\chi=U(\lambda;v,w)\chi=
\begin{pmatrix}v & -\lambda+w \\
1 & -v
\end{pmatrix}\chi.
\end{equation}
The resolution of the potentials $v,w$ in \eqref{eq:1.3} are not independent from the $a,b$ in \eqref{eq:1.2}, and the relations between them are the key
to the problem (see Section 2).

Another ingredient in our treatment is the methodology
of ``nonlinearisation" \cite{Cao1,Cao2}, which is related to the expansions of potentials $a,b,v,w$ in terms of squared eigenfunctions. In the present paper, following this method
we prove that the spectral problem (\ref{eq:1.3}) can be nonlinearised resulting in a finite
dimensional integrable Hamiltonian system which provides the essential conditions for constructing the relevant integrable symplectic map stemming from (\ref{eq:1.2}). Using this map, we
deduce several well-defined meromorphic functions on the spectral curve. Finally the lpKdV equation (\ref{eq:1.1}) is solved by solving the relevant Jacobi inversion problem in terms of theta functions on the hyperelliptic Riemann surface.
In contrasts to the usual cases treated in \cite{Cao,Cao3,Cao5}, where potentials themselves satisfy the
corresponding discrete models, the solutions here are expressed in terms of the derivative of a special theta function with respect to the auxiliary Darboux variable.

As it turns out, we conclude that one and the same discrete model can be solved through different Liouville integrable models. Inspired by this, we will also investigate the lattice potential modified KdV (lpmKdV) equation
\begin{equation}\label{eq:1.4}
\beta_{1}(\bar{u}\tilde{\bar{u}}-u\tilde{u})=\beta_{2}(\tilde{u}\tilde{\bar{u}}-u\bar{u}),
\end{equation}
which is closely related to the Hirota equation, i.e., the lattice sine-Gordon (lsG) equation whose algebro-geometric solutions have been discussed \cite{Hirota, Bobenko}.
In \cite{Cao3}, integrable symplectic maps and novel theta function solutions for equation (\ref{eq:1.4}) were constructed  through integrable Hamiltonian systems associated
with the continuous sG equation. In contract, in the present paper, we start from the Kaup-Newell spectral problem \cite{Cao4}
\begin{equation}\label{eq:1.5}
\partial_{x}\chi=V(\lambda;v,w)\chi=
\begin{pmatrix}\lambda^{2}/2 & \lambda v \\
\lambda w & -\lambda^{2}/2
\end{pmatrix}\chi,
\end{equation}
and the associated discrete spectral problem
\begin{equation}\label{eq:1.6}
\tilde{\chi}=(\lambda^{2}-\beta^{2})^{-1/2}D^{(\beta)}(\lambda;a)\chi,
\ \ D^{(\beta)}(\lambda;a)=
\begin{pmatrix}\lambda a & \beta \\
\beta & \lambda a^{-1}
\end{pmatrix}.
\end{equation}
In this example we actually have a different type of situation from the one of the previous examples \cite{Cao,Cao3,Cao5} since the relation between the discrete potential $a$ and the continuous potentials $v,w$ is implicit. However, based on the Lax structure of the Kaup-Newell equation, (\ref{eq:1.6}), the system can still be nonlinearised, in the sense mentioned above, as an integrable symplectic map.

In addition, we will investigate the lattice Schwarzian KdV (lSKdV) equation, first given in \cite{Capel},
\begin{equation}\label{eq:1.7}
\beta_{1}^{2}(\tilde{\bar{u}}-\tilde{u})(\bar{u}-u)=\beta_{2}^{2}(\tilde{\bar{u}}-\bar{u})(\tilde{u}-u),
\end{equation}
which expresses the cross-ratio of four points in the complex plane being equal to a constant. Equation (\ref{eq:1.7}) was used in \cite{Bobenko} to define a discrete
conformal map, consequently in \cite{Hertrich} solutions in terms of the Riemann theta function were written down in the context of the geometry of those conformal maps. Moreover, the finite genus solutions to lSKdV equation (\ref{eq:1.7}) are discussed, with the help of some finite dimensional integrable systems arising from one Lax matrix for the derivative Schwarzian KdV equation \cite{XCZ}. Interestingly, in \cite{NQC} by using the so called direct linearisation method, a more general form of (\ref{eq:1.7}) was deduced (nowadays sometimes referred to as NQC equation), which in different special parameter cases reduces to both (\ref{eq:1.1}), (\ref{eq:1.4}) and (\ref{eq:1.7}).
In the present case the relevant Hamiltonian system is associated with the continuous spectral problem
\begin{equation}\label{eq:1.8}
\partial_{x}\chi=W(\lambda;v,w)\chi=
\begin{pmatrix}-\lambda^{2}/2+v+w & \lambda v \\
-\lambda & \lambda^{2}/2-v-w
\end{pmatrix}\chi,
\end{equation}
which is curiously compatible with the dynamic problem of the Kac-van Moerbeke hierarchy \cite{Cao6}, and whose corresponding discrete spectral problem is given by
\begin{equation}\label{eq:1.9}
\tilde{\chi}=(\lambda^{2}-\beta^{2})^{-1/2}D^{(\beta)}(\lambda;a,s)\chi,
\ \ D^{(\beta)}(\lambda;a,s)=
\begin{pmatrix} \lambda a & \beta s\\
\beta s^{-1} & \lambda a^{-1}
\end{pmatrix}.
\end{equation}
Similarly, the parametrization of the potentials plays an essential role in this case.

The present paper is organised as follows. In Section 2, the construction of integrable symplectic maps and the resulting finite genus solutions to the lpKdV equation (\ref{eq:1.1}) are presented. In Section 3, we deal with the lpmKdV equation (\ref{eq:1.4}), and exploit the permutability of the integrable discrete phase
flows arising from the iteration of a novel parameter-family of integrable symplectic maps, leading to the corresponding finite-genus solutions to the partial difference equation. In Section 4, we study the lSKdV case and establish a useful relation between the two discrete potentials present in the same
spectral problem for the construction of the relevant integrable symplectic map. By this relation, a novel Lax pair for lSKdV equation (\ref{eq:1.7}) is obtained, different from the ones given in \cite{Capel,XCZ}. As a result, a recursion relation for the finite genus solutions is presented.

\section{ The lattice potential KdV equation}\setcounter{equation}{0}

In this section we will investigate the lpKdV equation (\ref{eq:1.1}) via integrable symplectic maps. For further calculations an integrable Hamiltonian system is needed, which provides the integrals, the spectral curve, etc.

\subsection {An integrable Hamiltonian system}

For the sake of self-containedness we first review some of the methodology of \cite{Cao}. This is based on the observation, going back to \cite{Flaschka}, that finite-dimensional integrable systems can be obtained by restricting the infinite-dimensional integrable systems to a finite-dimensional invariant manifold. To realise it, a good way is the nonlinearization of the spectral problem \cite{Cao1,Cao2}. Here we use this technique to construct a finite-dimensional integrable Hamiltonian system.

Consider $N$ copies of (\ref{eq:1.3}) with distinct eigenvalues $\alpha_1,\cdots,\alpha_N$, and write them in the vector form:
\begin{equation}\label{eq:2.1}
\begin{split}
&\partial_{x}p=vp-Aq+wq,\\
&\partial_{x}q=p-vq.
\end{split}
\end{equation}
where $A=\mathrm{diag}(\alpha_1,\cdots,\alpha_N)$. According to the principle of nonlinearisation \cite{Cao1,Cao2}, the
reflectionless potential can be expressed by the squared sum of eigenfunctions. In the present case, we shall impose some constraint on the potentials $v,w$, so that the linear equation (\ref{eq:2.1}) can be nonlinearised to produce a completely integrabe Hamiltonian system.

By \cite{Cao4,Cao6,Cao7}, we take the following Lax equation as a starting point:
\begin{equation}\label{eq:2.2}
\partial_xL(\lambda;p,q)=[U(\lambda;v,w),L(\lambda;p,q)],
\end{equation}
which is the compatibility condition of (\ref{eq:1.3}) and the eigenvalue problem $L(\lambda;p,q)\chi=\kappa \chi$.
The Lax matrix is computed in a similar way as in \cite{Cao4,Cao6,Cao7} from (\ref{eq:2.2}), and
adopts the following traceless form:
\begin{equation}\label{eq:2.3}
L(\lambda;p,q)=\begin{pmatrix} \sqrt{<q,q>}+Q_{\lambda}(p,q)& -\lambda-
Q_{\lambda}(p,p)
\\ 1+Q_{\lambda}(q,q)& -\sqrt{<q,q>}-Q_{\lambda}(p,q)\end{pmatrix},
\end{equation}
where $Q_{\lambda}(\xi,\eta)=<(\lambda I-A)^{-1}\xi,\eta>$, and $<\xi,\eta>=\Sigma^N_{j=1}\xi_j\eta_j$ is the usual inner product of two $N$-dimensional vectors $\xi,\eta$.

It is well-known that the characteristic polynomial $\mathrm{det}(\kappa I-L(\lambda;p,q))=\kappa^{2}+\mathrm{det}L(\lambda;p,q)$ represents the spectral curve of the system, which in turn gives rise to the integrals of the corresponding Hamiltonian system. In our case, we have the determinant
\begin{equation}\label{eq:2.4}
\begin{split}
\mathcal{F}_{\lambda}\overset{\triangle}{=}\mathrm{det}L(\lambda;p,q)=&\lambda+Q_{\lambda}(Aq,q)+Q_{\lambda}(p,p)-2\sqrt{<q,q>}Q_{\lambda}(p,q)\\
&+Q_{\lambda}(p,p)Q_{\lambda}(q,q)-Q_{\lambda}^{2}(p,q),
\end{split}
\end{equation}
where $Q_{\lambda}(Aq,q)=-<q,q>+\lambda Q_{\lambda}(q,q)$. Actually, $\mathcal{F}_{\lambda}$ is a rational function of $\lambda$, having simple poles at $\{\alpha_{j}\}_{j=1}^{N}$, since the coefficient of $(\lambda-\alpha_{j})^{-2}$ is zero. Thus,
\begin{equation}\label{eq:2.5}
\mathcal{F}_{\lambda}=\displaystyle \frac{\prod_{j=1}
^{N+1}(\lambda-\lambda_{j})}{\prod_{j=1}
^{N}(\lambda-\alpha_{j})}=\displaystyle\frac{\Lambda(\lambda)}{\alpha(\lambda)}.
\end{equation}
where $\Lambda(\lambda)=\prod_{j=1}
^{N+1}(\lambda-\lambda_{j}), \alpha(\lambda)=\Pi_{j=1}^{N}(\lambda-\alpha_{j})$. By virtue of general results of the theory of algebraic curves, cf. \cite{Farkas,Griffiths,Mumford}, the spectral curve associated with a 2-sheeted Riemann surface of genus $g=N$ is constructed according to the method elaborated in \cite{Moser,Moser1},
\begin{equation}\label{eq:2.6}
\mathcal {R}:\xi^{2}=-R(\lambda),
\end{equation}
where $R(\lambda)=\Lambda(\lambda)\alpha(\lambda)$. For values of $\lambda$ not corresponding to a branch point, there are two points $\mathfrak{p}(\lambda)$, $(\tau\mathfrak{p})(\lambda)$ on $\mathcal{R}$,
\begin{equation*}
\mathfrak{p}(\lambda)=\big(\lambda,\xi=\sqrt{-R(\lambda)}\big),\ \
(\tau\mathfrak{p})(\lambda)=(\lambda,\xi=-\sqrt{-R(\lambda)}\big).
\end{equation*}
with $\tau:\mathcal{R}\rightarrow\mathcal{R}$ the map of changing sheets.

Furthermore, from equation (\ref{eq:2.2}) it follows $\mathcal{F}_{\lambda}$ is independent of the argument $x$, therefore, $\mathcal{F}_{\lambda}$ can act as the generating function of the integrals associated with the Hamiltonian system (see below). In fact, setting
\begin{equation}\label{eq:2.7}
\mathcal{F}_{\lambda}=\lambda+\Sigma_{j=1}^{\infty}F_{j}\lambda^{-j},
\end{equation}
the coefficients in the expansion are given by
\begin{equation}\label{eq:2.8}
\begin{split}
F_{1}&=<Aq,q>+<p,p>-2\sqrt{<q,q>}<p,q>,\\
F_{l}&=<A^{l}q,q>+<A^{l-1}p,p>-2\sqrt{<q,q>}<A^{l-1}p,q>+\\
&\quad +\sum\limits_{j+k+2=l;j,k\geq0}\big(<A^{j}p,p><A^{k}q,q>-<A^{j}p,q><A^{k}p,q>\big), (l\geq 2).
\end{split}
\end{equation}

Inspired by the structure of the spectral curve (\ref{eq:2.6}), we find that $\mathcal{F}_{\lambda}$ is proportional to a perfect square of a quantity $ \mathcal{H}_{\lambda}$, namely
\begin{equation}\label{eq:2.9}
\lambda \mathcal{H}_{\lambda}^{2}=\mathcal{F}_{\lambda}.
\end{equation}
By using the expression (\ref{eq:2.7}), we obtain
\begin{equation}\label{eq:2.10}
\mathcal{H}_{\lambda}=1+\Sigma_{j=1}^{\infty}H_{j}\lambda^{-j-1},
\end{equation}
where
\begin{equation}\label{eq:2.11}
H_1(p,q)=\displaystyle\frac{1}{2}(<Aq,q>+<p,p>)-\sqrt{<q,q>}<p,q>.
\end{equation}
Now we consider $H_1(p,q)$ as a Hamiltonian function and calculate the corresponding Hamiltonian system. Fortunately, when choosing
\begin{equation}\label{eq:2.12}
(v,w)=(\sqrt{<q,q>},<p,q>/\sqrt{<q,q>}),
\end{equation}
in the reflectionless case, what the solution $(p_{j} ,q_{j})^{T}$ of the linear equation (\ref{eq:2.1}) satisfies, is
actually a system of nonlinear equations, which can be written in Hamiltonian form as:
\begin{align}\label{eq:2.13}
p_{x}=-\partial H_{1}/\partial q, \ \ q_{x}=\partial H_{1}/\partial p.
\end{align}

So far, we have finished the nonlinearisation of the eigenvalue problem (\ref{eq:1.3}), resulting in the Hamiltonian system (\ref{eq:2.13}). In the future analysis we need the following ingredients:

1) the canonical basis
$a_1,\cdots,a_g,b_1,\cdots,b_g$ of the homology group of contours.

2) the basis of the holomorphic differentials, written in the vector form
as
\begin{equation}\label{eq:2.14}
\vec{\omega}^\prime=(\omega_1^\prime,\cdots,\omega_g^\prime)^T, \ \
\omega^{\prime}_j=\lambda^{g-j}\text{d}\lambda/(2\xi),
\end{equation}
which can be normalized into $\vec{\omega}=C\vec{\omega}^\prime$, where
$C=(a_{jk})^{-1}_{g\times g}$, with $a_{jk}$ the integral of
$\omega_j^\prime$ along $a_k$ and $\vec{C}_{l}$ the \emph{l}-th column vector of $C$. Near the point at infinity, the following local expansion holds:
\begin{equation}\label{eq:2.15}
\vec{\omega}=[\vec{\Omega}_{1}+O(t^{2})]\mathrm{d}t,
\end{equation}
where $\vec{\Omega}_{1}=-\vec{C}_{1}$ and $t (t^{-2}=-\lambda)$ is the local coordinate for the branch point $\infty$.

3) the periodicity vectors $\vec\delta_k,\,\vec B_k$ defined as integrals of $\vec\omega$ along $a_k,\,b_k$, respectively. They span a lattice
$\mathscr T$ of periods, which defines the Jacobian variety $J(\mathcal R)=\mathbb C^g/\mathscr T$. The matrix $B =(\vec B_1,\ldots,\vec B_g)$ is used to construct the Riemann theta function
\begin{equation}\label{eq:2.16}
\theta(z,B)=\sum_{z^{\prime}\in \mathbb{Z}^{g}}\exp\pi\sqrt{-1}(<Bz^{\prime},z^{\prime}>+2<z,z^{\prime}>),\ \ z\in \mathbb{C}^{g}.
\end{equation}
The Abel map $\mathscr A: \text {Div}(\mathcal R)\rightarrow J(\mathcal R)$ is given as
\begin{equation}\label{eq:2.17}
\mathscr A(\mathfrak p)=\int_{\mathfrak{p}_0}^{\mathfrak{p}}\vec{\omega},\ \ \mathscr A(\Sigma n_{k}\mathfrak{p}_k)=\Sigma n_{k}\mathscr A(\mathfrak{p}_k).
\end{equation}
Thus by equation (\ref{eq:2.15}), we solve
\begin{equation}\label{eq:2.18}
-\mathscr A(\mathfrak p)=\int_{\mathfrak{p}}^{\mathfrak{p}_0}\vec{\omega}=\int_{\infty}^{\mathfrak{p}_0}\vec{\omega}+\int_{\mathfrak{p}}^{\infty}\vec{\omega}
=\eta-\vec{\Omega}_{1}t+O(t^{3}),\ \ \eta=\int_{\infty}^{\mathfrak{p}_0}\vec{\omega}.
\end{equation}

Now we discuss the complete integrability of (\ref{eq:2.13}) in the Liouville sense. Here we employ the $r$-matrix and the evolution of the Lax matrix along a certain phase flow, which can be used to encode the involution and independence of the integrals in our case.

Referring to \cite{Faddeev,Babelon,Gerdjikov}, we verify by direct computation that there are two matrix-valued functions, $r_{12}$ and $r_{21}$, on the symplectic space,
\begin{equation*}
r_{12}=
\begin{pmatrix}\displaystyle \frac{2}{\lambda-\mu}&0&\displaystyle\frac{-1}{\sqrt{<q,q>}}&0\\
0&0&\displaystyle \frac{2}{\lambda-\mu}&\displaystyle\frac{1}{\sqrt{<q,q>}}\\
0&\displaystyle \frac{2}{\lambda-\mu}&0&0\\0&0&0&\displaystyle
\frac{2}{\lambda-\mu}\end{pmatrix},
\end{equation*}
\begin{equation*}
 r_{21}=
\begin{pmatrix}\displaystyle \frac{2}{\mu-\lambda}&\displaystyle\frac{-1}{\sqrt{<q,q>}}&0&0\\
0&0&\displaystyle \frac{2}{\mu-\lambda}&0\\
0&\displaystyle
\frac{2}{\mu-\lambda}&0&\displaystyle\frac{1}{\sqrt{<q,q>}}\\
0&0&0&\displaystyle \frac{2}{\mu-\lambda}\end{pmatrix},
\end{equation*}
in terms of which the fundamental Poisson bracket between the Lax matrices takes the form:
\begin{equation}\label{eq:2.19}
\{L(\lambda)\underset{,}\otimes L(\mu)\} =
[r_{12},L_{1}(\lambda)]-[r_{21},L_{2}(\mu)],
\end{equation}
where $L(\lambda)$ is the abbreviation for $L(\lambda;p,q)$, $L_{1}(\lambda)=L(\lambda)\otimes I$, $L_{2}(\mu)=I\otimes
L(\mu)$ and $I$ is the usual unit matrix.

\noindent\textbf{Lemma 2.1.} The Lax matrix $L(\mu)$ satisfies the evolution equation along the $t_{\lambda}$-flow defined by the Hamiltonian vector field of $\mathcal{F}_{\lambda}$,
\begin{equation}\label{eq:2.20}
\mathrm{d}L(\mu)/\mathrm{d}t_{\lambda}=[W(\lambda,\mu),L(\mu)],
\end{equation}
where $W(\lambda,\mu)$ satisfies
\begin{align*}
&\displaystyle\frac{\mathrm{d}}{\mathrm{
d}t_{\lambda}}\begin{pmatrix}p_{j}\\q_{j}\end
{pmatrix}=\begin{pmatrix}-\partial \mathcal{F}_{\lambda}/\partial q_{j}\\
\partial \mathcal{F}_{\lambda}/\partial p_{j}\end
{pmatrix}=W(\lambda,\alpha_{j})\begin{pmatrix}p_{j}\\q_{j}\end
{pmatrix},\\
&W(\lambda,\mu)=\displaystyle\frac{2}{
\lambda-\mu}L(\lambda)+\displaystyle\frac{2L^{11}(\lambda)}{\sqrt{<q,q>}}\sigma_{+},\quad
\sigma_{+}=
\begin{pmatrix}0&1\\0&0\end{pmatrix}.
\end{align*}
\noindent \emph{Proof.} Since $L^{2}(\lambda)=-\mathcal{F}_{\lambda}I$, we obtain
\begin{align}\label{eq:2.21}
\begin{split}
\{L^{2}(\lambda)\underset{,}\otimes L(\mu)\}&=\{-\mathcal{F}_{\lambda}I\underset{,}\otimes L(\mu)\}\\
&=\begin{pmatrix}-\{\mathcal{F}_{\lambda},L(\mu)\}&0\\0&-\{\mathcal{F}_{\lambda},L(\mu)\}\end{pmatrix}\\
&=\begin{pmatrix}\mathrm{d}L(\mu)/\mathrm{d}t_{\lambda}&0\\0&\mathrm{d}L(\mu)/\mathrm{d}t_{\lambda}\end{pmatrix}.
\end{split}
\end{align}
By equation (\ref{eq:2.19}), we calculate the left hand side of (\ref{eq:2.21}) again and get
\begin{align}\label{eq:2.22}
\begin{split}
\{L^{2}(\lambda)\underset{,}\otimes L(\mu)\}=&L_{1}(\lambda)\{L(\lambda)\underset{,}\otimes L(\mu)\} +\{L(\lambda)\underset{,}\otimes L(\mu)\}L_{1}(\lambda)\\
=&-L_{1}(\lambda)r_{21}L_{2}(\mu)+L_{1}(\lambda)L_{2}(\mu)r_{21}\\
&-r_{21}L_{2}(\mu)L_{1}(\lambda)+L_{2}(\mu)r_{21}L_{1}(\lambda)\\
=&-[L_{1}(\lambda)r_{21}+r_{21}L_{1}(\lambda),L_{2}(\mu)]\\
=&\begin{pmatrix}[W(\lambda,\mu),L(\mu)]&0\\0&[W(\lambda,\mu),L(\mu)]\end{pmatrix},
\end{split}
\end{align}
where we use the formulas $L_{1}^{2}(\lambda)=-\mathcal{F}_{\lambda}I$ and $L_{1}(\lambda)L_{2}(\mu)=L_{2}(\mu)L_{1}(\lambda)=L(\lambda)\otimes L(\mu)$ which are easily got by some calculations. Then comparing (\ref{eq:2.21}) with (\ref{eq:2.22}), equation (\ref{eq:2.20}) is verified. \hfill $\Box$

\noindent As a corollary of Lemma 2.1, we have $\mathrm{d}L^{2}(\mu)/\mathrm{d}t_{\lambda}=[W(\lambda,\mu),L^{2}(\mu)]$, which implies
\begin{equation}\label{eq:2.23}
\mathrm{d}\mathcal{F}_{\mu}/\mathrm{d}t_{\lambda}=\{\mathcal{F}_{\mu}, \mathcal{F}_{\lambda}\} = 0, \ \ \forall \mu, \lambda \in\mathbb{C},
\end{equation}
by using the formula $L^{2}(\mu)=-\mathcal{F}_{\mu}I$ and the fact that $\mathcal{F}_{\lambda}$ is Hamiltonian for the $t_{\lambda}$-flow. As a consequence we have
\begin{align*}
\begin{split}
&\{\mathcal{F}_{\mu},\mathcal{F}_{\lambda}\}=\{\mathcal{F}_{\mu},\mathcal{H}_{\lambda}\}=\{\mathcal{H}_{\mu},\mathcal{H}_{\lambda}\}=0,\quad \forall \mu, \lambda \in\mathbb{C},\\
&\{F_{j},F_{k}\}=\{F_{j},H_{k}\}=\{H_{j},H_{k}\}=0,\quad \forall j,k=1,2,3,\ldots\ \ ,
\end{split}
\end{align*}
implying that $F_{1},\ldots,F_{N}$ are in pairwise involution, moreover, they are integrals for Hamiltonian system (\ref{eq:2.13}).

In the theory of Liouville integrability the functional independence of $F_{1},\ldots,F_{N}$ plays a fundamental role \cite{Arnold0,Arnold,Abraham}. In order to prove the latter, we introduce the elliptic variables $\nu_{j}$, i.e., curvilinear orthogonal coordinates \cite{Lame},
\begin{equation}\label{eq:2.24}
L^{21}(\lambda)=1+Q_{\lambda}(q,q)=\prod_{j=1}^{N}\displaystyle\frac{
\lambda-\nu_{j}}{ \lambda-\alpha_{j}}=\displaystyle\frac{
\mathfrak{n}(\lambda)}{ \alpha(\lambda)}.
\end{equation}
A resolution of (\ref{eq:2.24}) is given in terms of the quasi-Abel-Jacobi variable $\vec\phi^\prime$ and the Abel-Jacobi variable $\vec\phi$, which are defined as
\begin{equation}\label{eq:2.25}
\vec\phi^\prime=\sum_{k=1}^g\int_{\mathfrak p_0}^{\mathfrak
p(\nu_k)}\vec\omega^\prime,\quad\vec\phi=C\vec\phi^\prime=\mathscr
{A}(\sum_{k=1}^g\mathfrak p(\nu_k)),
\end{equation}
taking values in the Jacobian variety $J(\mathcal R)$, by using the Abel map $\mathscr{A}$.

Consider one of the entries of matrix equation (\ref{eq:2.20}):
\begin{equation}\label{eq:2.26}
\mathrm{d}L^{21}(\mu)/\mathrm{d}t_{\lambda}=2(W^{21}(\lambda,\mu)L^{11}(\mu)-W^{11}(\lambda,\mu)L^{21}(\mu)),
\end{equation}
Since $\mathcal{F}_{\lambda}=-(L^{11}(\lambda))^{2}-L^{12}(\lambda)L^{21}(\lambda)$, we find
\begin{equation*}
L^{11}(\nu_{k})=\sqrt{-R(\nu_{k})}/\alpha(\nu_{k}),
\end{equation*}
as a consequence of equation (\ref{eq:2.5}). Evaluating equation (\ref{eq:2.26}) at the point $\mu=\nu_{k}$, we obtain the evolution
of the elliptic variables $\nu_{k}$ along the $t_{\lambda}$-flow,
\begin{equation*}
\displaystyle\frac{1}{2\sqrt{-R(\nu_{k})}}\displaystyle\frac{
\mathrm{d}\nu_{k}}{
\mathrm{d}t_{\lambda}}=\frac{\displaystyle-2}{\displaystyle\alpha(\lambda)}\displaystyle\frac{
\mathfrak{n}(\lambda)}{
(\lambda-\nu_{k})\mathfrak{n}^{\prime}(\nu_{k})},\quad(1\leq
k\leq g),
\end{equation*}
which are the Dubrovin equations for our case \cite{Gesztesy,Dubrovin1,Nijhoff2}. Then by means of the Lagrange interpolation formula for polynomials, we have
\begin{equation*}
\sum\limits
_{k=1}^{g}\displaystyle\frac{
\nu_{k}^{g-l}}{2\sqrt{-R(\nu_{k})}}\displaystyle\frac{\mathrm{d}\nu_{k}}{
\mathrm{d}t_{\lambda}}=\displaystyle\frac{-2}{\alpha(\lambda)}\sum\limits
_{k=1}^{g}\displaystyle\frac{\nu_{k}^{g-l}\mathfrak{n}(\lambda)}{(\lambda-\nu_{k})\mathfrak{n}^{\prime}(\nu_{k})}
=\displaystyle\frac{-2}{\alpha(\lambda)}\lambda^{g-l},\quad(1\leq
l\leq g),
\end{equation*}
which can be rewritten in a simple form
\begin{equation}\label{eq:2.27}
\{\phi_{l}^{'},\mathcal{F}_{\lambda}\}=\displaystyle\frac{
\mathrm{d}\phi_{l}^{'}}{\mathrm{d}t_{\lambda}}=\displaystyle\frac{-2}{\alpha(\lambda)}\lambda^{g-l},\quad(1\leq
l\leq g),
\end{equation}
using the quasi-Abel-Jacobi variable $\vec\phi^{\prime}=(\phi_1^\prime,\cdots,\phi_g^\prime)^{T}$ given by (\ref{eq:2.25}). Expanding both sides of equation (\ref{eq:2.27}), we obtain
\begin{equation*}
\sum\limits_{j=1}^{\infty}\{\phi_{l}^{'},F_{j}\}\lambda^{-j}=
\displaystyle\frac{-2\lambda^{-l}}{\Pi_{k=1}^{g}(1-\alpha_{k}\lambda^{-1})}
=-2\sum\limits_{i=0}^{\infty}A_{i}\lambda^{-(i+l)}=-2\sum\limits_{j=-\infty}^{\infty}A_{j-l}\lambda^{-j},
\end{equation*}
with $A_{0}=1, A_{-l}=0, \forall l\in \mathbb {N}$. Hence, we get
\begin{equation}\label{eq:2.28}
\displaystyle\frac{
\partial(\phi_{1}^{'},\ldots,\phi_{g}^{'})}{ \partial
(t_{1},\ldots,t_{g})}=\big(\{\phi_{l}^{'},F_{j}\}\big)_{g\times g}=
-2\begin{pmatrix}1&A_{1}&A_{2}&\ldots&A_{g-1}\\ \quad
&1&A_{1}&\ldots&A_{g-2}\\ \quad & \quad &\ddots&\ddots&\vdots\\
\quad& \quad & \quad&\ddots&A_{1}\\ \quad & \quad
&\quad&\quad&1\end{pmatrix},
\end{equation}
where $t_{j}$ is the flow variable, i.e., $\mathrm{d}G/\mathrm{d}t_{j}=\{G, F_{j}\}$ for any smooth function $G(p,q)$. We note in passing that the matrix (\ref{eq:2.28}) is non-degenerate, hence $\mathrm{d}F_{1},\ldots,\mathrm{d}F_{N}$ are linearly independent throughout each cotangent space $T_{y}^{*}\mathbb{R}^{2N}, \forall y\in \mathbb{R}^{2N}$. Indeed suppose $\Sigma_{j=1}^{N}c_{j}\mathrm{d}F_{j}=0$, then
$\Sigma_{j=1}^{N}c_{j}\{\phi_{l}^{'},F_{j}\}=0,\forall l $, which implies that $c_{j}=0,\forall j$. Thus we arrive at the main result:

\noindent \textbf{Proposition 2.1.} The Hamiltonian system (\ref{eq:2.13}) is completely integrable in the sense of Liouville, possessing the integrals
$F_{1},\ldots,F_{N}$, which are in involution w.r.t. the canonical Poisson brackets and functionally independent on $\mathcal{N}=(\mathbb{R}^{2N},\mathrm{d}p\wedge \mathrm{d}q)$.

\subsection {An integrable symplectic map}

 We now use the results in Section 2.1 to construct an integrable symplectic map. Motivated by \cite{Cao}, we define the following linear map on $\mathcal{N}$,
\begin{equation}\label{eq:2.29}
\begin{array}{lcl}
S_{\beta}:
\left(\begin{array}{c}
\tilde{p}_{j}\\
\tilde{q}_{j}
\end{array}\right)=
(\alpha_{j}-\beta)^{-1/2}D^{(\beta)}(\alpha_{j};a,b)
\left(\begin{array}{c}
 p_{j}\\
 q_{j}
 \end{array}\right), \ \ (1\leq j\leq N),
\end{array}
\end{equation}
where $D^{(\beta)}$ is the relevant Darboux matrix given in (\ref{eq:1.2}). Similarly, we will find the constraint on the discrete potentials $a,b$, under which $S_{\beta}$ can be nonlinearised to derive an integrable symplectic correspondence. This can be seen to arise from the following discrete Lax equation:
\begin{equation}\label{eq:2.30}
\mathcal {D}\overset{\triangle}{=}L(\lambda;\tilde{p},\tilde{q})D^{(\beta)}(\lambda;a,b)-D^{(\beta)}(\lambda;a,b)L(\lambda;p,q)=0.
\end{equation}
In fact, by direct calculations, we have
\begin{align*}
\mathcal {D}=&
\left(\begin{array}{cc} \tilde{v }& -\lambda\\
1 & -\tilde{v}
\end{array}\right)D^{(\beta)}(\lambda)-D^{(\beta)}(\lambda)
\left(\begin{array}{cc} v & -\lambda\\
1 & -v
\end{array}\right)\\
&+\sum\limits_{j=1}^{N}
\frac{\displaystyle{1}}{\displaystyle{\lambda-\alpha_{j}}}
(\tilde{\varepsilon}_{j}D^{(\beta)}(\lambda)-D^{(\beta)}(\lambda)\varepsilon_{j}),
\end{align*}
where $v$ is given by equation (\ref{eq:2.12}) and
\begin{equation*}
\varepsilon_{j}=\begin{pmatrix}p_{j}q_{j} & -p_{j}^{2}\\q_{j}^{2} &
-p_{j}q_{j}\end{pmatrix}
\end{equation*}
satisfies $\tilde{\varepsilon}_{j}D^{(\beta)}(\alpha_{j};a,b)-D^{(\beta)}(\alpha_{j};a,b)\varepsilon_{j}=0$. Then the entries of the matrix $\mathcal {D}$ are expressed as
\begin{align*}
\mathcal {D}^{11}&=a(\tilde{v}-v-b)+v^{2}-\beta,\\
\mathcal {D}^{12}&=-\lambda (b-a+v+\tilde{v})+(\beta+ab)(v+\tilde{v})-<\tilde{p},\tilde{q}>-<p,q>,\\
\mathcal {D}^{21}&=a-\tilde{v}-v-b,\\
\mathcal {D}^{22}&=b(a-\tilde{v}+v)-v^{2}+\beta.
\end{align*}
Hence, from the formula for $\mathcal {D}^{21}$ we choose the restriction
\begin{equation}\label{eq:2.31}
a=b+v+\tilde{v},
\end{equation}
Substituting (\ref{eq:2.31}) into the other components, we obtain
\begin{align*}
\mathcal {D}^{12}&=b\mathcal {D}^{11}+a\mathcal {D}^{22},\\
\mathcal {D}^{11}&=-\mathcal {D}^{22}=-P^{(\beta)}(b;p,q),
\end{align*}
where
\begin{equation}\label{eq:2.32}
P^{(\beta)}(b;p,q)=L^{21}(\beta)b^{2}+2L^{11}(\beta)b-L^{12}(\beta),
\end{equation}
by using
\begin{align*}
\tilde{p}_{j}&=(\alpha_{j}-\beta)^{-1/2}[ap_{j}+(-\alpha_{j}+\beta+ab)q_{j}],\\
\tilde{q}_{j}&=(\alpha_{j}-\beta)^{-1/2}(p_{j}+bq_{j}),
\end{align*}
derived from (\ref{eq:2.29}). Therefore, we assert that the roots of the quadratic equation $P^{(\beta)}(b;p,q)=0$ give rise to an explicit constraint on $b$,
\begin{equation}\label{eq:2.33}
b=f_{\beta}^{2}(p,q)=\displaystyle
\frac{1}{1+Q_{\beta}(q,q)}(-\sqrt{<q,q>}-Q_{\beta}(p,q)\pm\frac{\sqrt{-R(\beta)}}{\alpha(\beta)}),
\end{equation}
Actually they are the values of a meromorphic function on the curve $\mathcal{R}$ defined by (\ref{eq:2.6}),
\begin{equation*}
\mathfrak{B}(\mathfrak{p})=\displaystyle
\frac{1}{1+Q_{\beta}(q,q)}(-\sqrt{<q,q>}-Q_{\beta}(p,q)+\frac{\xi}{\alpha(\beta)}),
\end{equation*}
at the points $\mathfrak{p}(\beta)$ and $(\tau\mathfrak{p})(\beta)$, respectively. Then by the relation (\ref{eq:2.31}), we get
\begin{equation}\label{eq:2.34}
a=f_{\beta}^{1}(p,q)=f_{\beta}^{2}(p,q)+\sqrt{<\tilde{q},\tilde{q}>}+\sqrt{<q,q>}.
\end{equation}
Though doubled-valued as functions of $\beta\in \mathbb{C}$, (\ref{eq:2.33}) and (\ref{eq:2.34}) are single-valued as functions of $\mathfrak{p}(\beta)\in \mathcal{R}$. Thus we get the following expression for the discrete potentials in term of the square of eigenfunctions:
\begin{equation}\label{eq:2.35}
(a,b)=f_{\beta}(p,q)=(f_{\beta}^{1}(p,q),f_{\beta}^{2}(p,q)),
\end{equation}
by which the linear map $S_{\beta}$ given by (\ref{eq:2.29}) becomes a nonlinear map
\begin{equation}\label{eq:2.36}
S_{\beta}: \ \ \begin{pmatrix}\tilde{p} \\ \tilde{q}\end{pmatrix}=(A-\beta)^{-1/2}
\begin{pmatrix}ap+(-A+\beta+ab)q\\p+bq\end{pmatrix}\Bigg|_{(a,b)=f_{\beta}(p,q)}.
\end{equation}
Here we use the same symbol $S_{\beta}$ for short.

\noindent \textbf{Proposition 2.2.} The above map (\ref{eq:2.36}) is an integrable symplectic map, under which the quantities $F_{1},\ldots,F_{N}$ on phase space given by (\ref{eq:2.8}) are invariant.

\noindent \emph{Proof.} According to the above analysis, we have
\begin{equation}\label{eq:2.37}
L(\lambda;\tilde{p},\tilde{q})D^{(\beta)}\big(\lambda;f_{\beta}(p,q)\big)
-D^{(\beta)}\big(\lambda;f_{\beta}(p,q)\big)L(\lambda;p,q)=0,
\end{equation}
by substituting (\ref{eq:2.35}) into the left hand side of the discrete Lax equation (\ref{eq:2.30}). Taking the determinant on (\ref{eq:2.37}), we obtain $\tilde{\mathcal{F}}_{\lambda}=\mathcal{F}_{\lambda}$ which implies $F_{j}(\tilde{p},\tilde{q})=F_{j}(p,q)$, i.e., $S_{\beta}^{*}F_{j}=F_{j},1\leq j\leq N$.

\noindent In order to get the symplectic property of (\ref{eq:2.36}), we calculate
\begin{eqnarray}\label{eq:2.38}
\sum\limits_{j=1}^{N} (\mathrm{d}\tilde{p}_{j} \wedge
\mathrm{d}\tilde{q}_{j}-\mathrm{d}p_{j} \wedge\mathrm{d}
q_{j})=-\displaystyle\frac{1}{2}\mathrm{d}P^{(\beta)}(b;p,q) \wedge
\mathrm{d}\sqrt{<\tilde{q},\tilde{q}>},
\end{eqnarray}
where $P^{(\beta)}(b;p,q)$ is given by (\ref{eq:2.32}). Thus $S_{\beta}^{*}(\mathrm{d}p\wedge \mathrm{d}q)=\mathrm{d}p\wedge \mathrm{d}q$ under the constraint (\ref{eq:2.35}). \hfill $\Box$

Proposition 2.2 implies the Liouville integrability of the symplectic map \cite{Quispel,Bruschi,Suris,Veselov}. Considering the map (\ref{eq:2.36}) as an iterative map, we can create a discrete orbit starting from an initial data $(p_0,q_0)\in \mathbb{R}^{2N}$. Thus we are able to define a discrete phase flow $\big(p(m),q(m)\big)=S^{m}_{\beta}(p_{0},q_{0})$ repeated application of the map, i.e., $ S^{m}_{\beta}=S_{\beta}\circ S^{m-1}_{\beta}$. The potentials along the $S^{m}_{\beta}$-flow are,
\begin{eqnarray}\label{eq:2.39}
&&(a(m),b(m))=(a_{m},b_{m}) = f_{\beta}\big(p(m),q(m)\big) =
f_{\beta}\big(S^{m}_{\beta}(p_{0},q_{0})\big),\\\label{eq:2.40}
&&a_{m}=b_{m}+v_{m}+v_{m+1},\ \ \mathrm{or}  \ \ a=b+v+\tilde{v},
\end{eqnarray}
 where $v_{m} = \sqrt{<q(m),q(m)>}$. Then equation (\ref{eq:2.37}) can be written in the form
\begin{equation}\label{eq:2.41}
L_{m+1}(\lambda)D^{(\beta)}_{m}(\lambda) =
D^{(\beta)}_{m}(\lambda)L_{m}(\lambda),
\end{equation}
where we have used the abbreviations $L_{m}(\lambda)
=L(\lambda;p(m),q(m)), D^{(\beta)}_{m}(\lambda) =
D^{(\beta)}(\lambda;a_{m},b_{m})$.

\noindent We remark that since $\mathcal{F}_{\lambda}\big(S_{\beta}(p,q)\big)=\mathcal{F}_{\lambda}(\tilde{p},\tilde{q})=\mathcal{F}_{\lambda}(p,q)$, we have $\mathcal{F}_{\lambda}\big(p(m),q(m)\big)=\mathcal{F}_{\lambda}(p_{0},q_{0})$. Thus the spectral curve $\mathcal {R}$ is invariant under the $S^{m}_{\beta}$-flow.

\subsection{ The finite genus solutions to the lpKdV equation}

 Our aim is to calculate finite gap classes of exact solutions for lpKdV equation (\ref{eq:1.1}). We now consider $D^{(\beta)}_{m}(\lambda)$ as a difference operator and $L_{m}(\lambda)$ as an algebra operator. The above commutativity relation (\ref{eq:2.41}) between them reminds us of the Burchnall-Chaundy theory \cite{Burchnall}, for commutative differential operators, the discrete analogue of which we formulate below \cite{Cao}, and the Baker-Akhiezer functions \cite{Baker,Akhiezer} as well. Therefore, we introduce a new discrete spectral problem with potentials $a_{m}, b_{m}$ as
\begin{equation}\label{eq:2.42}
h_{\beta}(m+1,\lambda) =
D^{(\beta)}_{m}(\lambda)h_{\beta}(m,\lambda),
\end{equation}
and investigate its fundamental solution matrix $M_{\beta}(m,\lambda)$ with $M_{\beta}(0,\lambda) = I$. Fortunately, the solution space $\varepsilon_{\lambda}$ of equation \eqref{eq:2.42} is invariant under the action of the linear operator $L_{m}(\lambda)$. In fact, if $h\in \varepsilon_{\lambda}$, then by \eqref{eq:2.41},
\begin{equation*}
(Lh)_{m+1}=L_{m+1}(D^{(\beta)}_{m}h_{m})=D^{(\beta)}_{m}(Lh)_{m},
\end{equation*}
which implies $Lh\in \varepsilon_{\lambda}$.

In the invariant space $\varepsilon_{\lambda}$, the linear operator $L_{m}(\lambda)$ has two eigenvalues $\rho^{\pm}_{\lambda}$, independent of the
discrete argument $m$ due to Proposition 2.2,
\begin{equation}\label{eq:2.43}
\rho^{\pm}_{\lambda} = \pm\rho_{\lambda} =
\pm\displaystyle\sqrt{-\mathcal{F}_{\lambda}} =
\pm\sqrt{-R(\lambda)}/\alpha(\lambda),
\end{equation}
which are the values of a well-defined meromorphic function  $\xi(\mathfrak{p})/\alpha\big(\lambda(\mathfrak{p})\big)$ on
$\mathcal {R}$ at the points $\mathfrak{p}(\lambda), (\tau\mathfrak{p})(\lambda)$ respectively.

\noindent The corresponding eigenvectors $h_{\beta,\pm}$ are given by
\begin{equation}\label{eq:2.44}
 \big(L_{m}(\lambda)-\rho^{\pm}_{\lambda}\big)h_{\beta,\pm}(m,\lambda)= 0.
\end{equation}
Simultaneously $h_{\beta,\pm}$ are solutions of the equation \eqref{eq:2.42}, thus we have
\begin{eqnarray}\label{eq:2.45}
h_{\beta,\pm}(m+1,\lambda) =
D^{(\beta)}_{m}(\lambda)h_{\beta,\pm}(m,\lambda),
\end{eqnarray}
and in this sense the $h_{\beta,\pm}$ are common eigenvectors for $D^{(\beta)}_{m}(\lambda)$ and $L_{m}(\lambda)$. Since the rank of $L_{m}(\lambda)\mp \sqrt{-\mathcal{F}_{\lambda}}$ is equal to 1, in each case the common eigenvector is uniquely determined up to a constant factor. We choose
\begin{equation}\label{eq:2.46}
h_{\beta,\pm}(m,\lambda) =
\begin{pmatrix}h_{\beta,\pm}^{(1)}(m,\lambda)\\h_{\beta,\pm}^{(2)}(m,\lambda)\end{pmatrix}
 = M_{\beta}(m,\lambda)
 \begin{pmatrix}c_{\lambda}^{\pm}\\1 \end{pmatrix}.
\end{equation}
By letting $m=0$ in equations \eqref{eq:2.44} and \eqref{eq:2.46}, we solve
\begin{equation}\label{eq:2.47}
c^{\pm}_{\lambda} = \displaystyle\frac{L^{11}_{0}(\lambda) \pm
\rho_{\lambda}}{L^{21}_{0}(\lambda)} =
-\frac{\displaystyle{L^{12}_{0}(\lambda)}}{\displaystyle{L^{11}_{0}(\lambda)\mp\rho_{\lambda}}}.
\end{equation}
Hence $c^{+}_{\lambda}c^{-}_{\lambda} =-L^{12}_{0}(\lambda)/L^{21}_{0}(\lambda)$, and $c^{+}_{\lambda}, c^{-}_{\lambda}$ are two branches of a meromorphic function on two sheets of $\mathcal {R}$, since $L_{0}^{jk}, j,k=1,2$ are rational functions of $\lambda$ apart from $\rho_{\lambda}$.

Following \cite{Baker,Akhiezer,Krichever,Cao}, we now investigate a well-defined meromorphic function $\mathfrak{h}_{\beta}^{(2)}(m,\mathfrak{p})$ on $\mathcal{R}$, i.e., Baker-Akhiezer function, with values $h^{(2)}_{\beta,+}(m,\lambda)$ and $h^{(2)}_{\beta,-}(m,\lambda)$ at the points $\mathfrak{p}(\lambda)$ and $(\tau\mathfrak{p})(\lambda)$ respectively. It turns out that the explicit expression of $\mathfrak{h}_{\beta}^{(2)}(m,\mathfrak{p})$ is the key to the solutions for lpKdV equation (\ref{eq:1.1}), in terms of theta functions. According to the theory of Riemann surfaces \cite{Farkas,Griffiths,Mumford}, we need to consider the zeros and poles of $\mathfrak{h}_{\beta}^{(2)}(m,\mathfrak{p})$.

\noindent{\textbf{Lemma 2.2.}} The following formula holds \cite{Cao}:
\begin{equation}\label{eq:2.48}
h^{(2)}_{\beta,+}(m,\lambda) \cdot h^{(2)}_{\beta,-}(m,\lambda) =
(\lambda-\beta)^{m}\prod\limits_{j=1}^{g}
\displaystyle\frac{\lambda-\nu_{j}(m)}{\lambda-\nu_{j}(0)}.
\end{equation}
\noindent\emph{Proof.} Since $M_{\beta}(m,\lambda)$ is the solution matrix of the equation \eqref{eq:2.42}, we have
\begin{equation}\label{eq:2.49}
M_{\beta}(m+1,\lambda) =
D^{(\beta)}_{m}(\lambda)M_{\beta}(m,\lambda),
\end{equation}
and by induction, we obtain
\begin{equation}\label{eq:2.50}
M_{\beta}(m,\lambda) =
D^{(\beta)}_{m-1}(\lambda)D^{(\beta)}_{m-2}(\lambda)\ldots
D^{(\beta)}_{0}(\lambda).
\end{equation}
Then from the commutativity relation \eqref{eq:2.41}, the action of the algebra operator $L_{m}(\lambda)$ on $M_{\beta}(m,\lambda)$ gives rise to
\begin{equation}\label{eq:2.51}
L_{m}(\lambda)M_{\beta}(m,\lambda) =
M_{\beta}(m,\lambda)L_{0}(\lambda).
\end{equation}
Finally we calculate,
\begin{align*}
\begin{split}
\begin{pmatrix} h^{(1)}_{\beta,+}h^{(1)}_{\beta,-} &
h^{(1)}_{\beta,+}h^{(2)}_{\beta,-}\\h^{(2)}_{\beta,+}h^{(1)}_{\beta,-}
& h^{(2)}_{\beta,+}h^{(2)}_{\beta,-}\end {pmatrix}& =
M_{\beta}(m,\lambda)
\begin{pmatrix} c^{+}_{\lambda}c^{-}_{\lambda} & c^{+}_{\lambda}\\
c^{-}_{\lambda} & 1 \end {pmatrix}M_{\beta}^{T}(m,\lambda)\\
&=\displaystyle\frac{1}{L^{21}_{0}(\lambda)}M_{\beta}(m,\lambda)
[L_{0}(\lambda) + \rho_{\lambda}]\mathrm{i}\sigma_{2}M_{\beta}^{T}(m,\lambda)\\
&=\displaystyle\frac{1}{L^{21}_{0}(\lambda)}[L_{m}(\lambda)
+ \rho_{\lambda}]M_{\beta}(m,\lambda)\mathrm{i}\sigma_{2}M_{\beta}^{T}(m,\lambda)\\
&=\displaystyle\frac{1}{L^{21}_{0}(\lambda)}[L_{m}(\lambda) +
\rho_{\lambda}]\mathrm{i}\sigma_{2}(\lambda-\beta)^{m},
\end{split}
\end{align*}
where $\sigma_{2}$ is the Pauli matrix. Thus,
$h^{(2)}_{\beta,+}h^{(2)}_{\beta,-} =
(\lambda-\beta)^{m}L^{21}_{m}(\lambda)/L^{21}_{0}(\lambda),$ which
implies \eqref{eq:2.48} by using \eqref{eq:2.24}. \hfill $\Box$

Lemma 2.2 gives the total zeros and some poles. We now exhibit the remaining poles stemming from the asymptotic behaviors \cite{Chen}.

\noindent{\textbf{Lemma 2.3.}} In the neighborhood of $\infty$, the following formula reads:
\begin{equation}\label{eq:2.52}
h_{\beta,\pm}^{(2)}(m,\lambda)=(\pm t)^{-m}[1+O(t)].
\end{equation}
\noindent\emph{Proof.} By using equation \eqref{eq:2.46}, we have
\begin{equation}\label{eq:2.53}
h^{(2)}_{\beta,\pm}(m,\lambda)=c_{\lambda}^{\pm}M_{\beta}^{21}(m,\lambda)+M_{\beta}^{22}(m,\lambda).
\end{equation}
Thus the asymptotic behaviors for $c_{\lambda}^{\pm}$ and $M_{\beta}(m,\lambda)$ are needed. From equation \eqref{eq:2.47}, we solve
\begin{equation}\label{eq:2.54}
c^{\pm}_{\lambda} = \pm\sqrt{-\lambda}[1 + O(\lambda^{-1/2})], \ \ (\lambda\rightarrow \infty).
\end{equation}
Moreover, as $\lambda\rightarrow \infty$, we get
\begin{align}\label{eq:2.55}
\begin{split}
&M_{\beta}(2k,\lambda) = \begin{pmatrix}(-\lambda)^{k}[1 +
O(\lambda^{-1})] & O(\lambda^{k})\\
O(\lambda^{k-1}) & (-\lambda)^{k}[1 +
O(\lambda^{-1})]\end{pmatrix},\\
& M_{\beta}(2k+1,\lambda) =
\begin{pmatrix}O(\lambda^{k}) & (-\lambda)^{k+1}[1 +
O(\lambda^{-1})]\\(-\lambda)^{k}[1 + O(\lambda^{-1})] &
O(\lambda^{k})\end {pmatrix},
\end{split}
\end{align}
by equation \eqref{eq:2.50} and induction.

\noindent Substituting \eqref{eq:2.54} and \eqref{eq:2.55} into \eqref{eq:2.53}, we obtain
\begin{align*}
\begin{split}
&h_{\beta,\pm}^{(2)}(2k,\lambda)=O(\lambda^{k-1/2})+(-\lambda)^{k}[1+O(\lambda^{-1})]=(\pm t)^{-2k}[1+O(t)],\\
&h_{\beta,\pm}^{(2)}(2k+1,\lambda)=\pm(-\lambda)^{k+1/2}[1+O(\lambda^{-1/2})]=(\pm t)^{-2k-1}[1+O(t)],
\end{split}
\end{align*}
whose unified form is exactly equation \eqref{eq:2.52}. \hfill $\Box$

The spectral curve $\mathcal {R}$ has a local coordinate $t=(-\lambda)^{-1/2}$ at the branch points $\infty$. Thus, by equation \eqref{eq:2.52},
$\mathfrak{h}_{\beta}^{(2)}(m,\mathfrak{p})$ has a pole at $\infty$ of order $m$. Considering zeros and other poles by equation \eqref{eq:2.48}, we arrive at

\noindent{\textbf{Proposition 2.3.}} The Baker-Akhiezer function $\mathfrak{h}_{\beta}^{(2)}(m,\mathfrak{p})$ has divisors
\begin{equation}\label{eq:2.56}
\mathrm{Div}(\mathfrak{h}_{\beta}^{(2)}(m,\mathfrak{p}))=\sum_{j=1}^{g}\big(\mathfrak{p}(\nu_{j}(m))-\mathfrak{p}(\nu_{j}(0))\big)
+ m\big(\mathfrak{p}(\beta)-{\infty}\big).
\end{equation}
According to \cite{Farkas,Griffiths,Mumford}, for any two distinct points
$\mathfrak{q},\mathfrak{r}\in\mathcal{R}$, there exists a dipole
$\omega[\mathfrak{q},\mathfrak{r}]$, an Abel differential of the
third kind, with residues $1,-1$ at the poles
$\mathfrak{q},\mathfrak{r}$, respectively, satisfying
\begin{equation*}
\int_{a_j}\omega[\mathfrak{q},\mathfrak{r}]=0,\quad\int_{b_j}\omega[\mathfrak{q},\mathfrak{r}]=\int_{\mathfrak{r}}^{\mathfrak{q}}\omega_j,
\quad(j=1,\cdots,g).
\end{equation*}
Decomposing the meromorphic differential as the combination
\begin{equation*}
\mathrm{d}\mathrm{ln} \mathfrak{h}_{\beta}^{(2)}(m,\mathfrak{p})=\sum_{j=1}^{g}\omega[\mathfrak{p}(\nu_{j}(m)),\mathfrak{p}(\nu_{j}(0))]+m\omega[\mathfrak{p}(\beta),\infty]
+\sum_{j=1}^{g}\gamma_{j}\omega_{j}+\Omega,
\end{equation*}
where $\gamma_{j}$ are constants, and $\Omega$ is the Abelian differential of the second kind,
with residues equal to zero at all poles. Refer to \cite{Toda}, the differential leads to
\begin{equation}\label{eq:2.57}
\sum_{j=1}^{g}\int_{\mathfrak{p}(\nu_{j}(0))}^{\mathfrak{p}(\nu_{j}(m))}
\vec{\omega} +
m\displaystyle\int_{\infty}^{\mathfrak{p}(\beta)}\vec{\omega} \equiv
0, \quad (\mathrm{mod}\mathscr T).
\end{equation}
Now we consider the formula \eqref{eq:2.57} from the Abel-Jacobi variables, and find that the $S^{m}_{\beta}$-flow viewed in the Jacobian variety $J(\mathcal R)$ is linear, i.e.,
\begin{equation}\label{eq:2.58}
\vec{\phi}(m)\equiv\vec{\phi}(0)+m\vec\Omega_{\beta},\quad\quad(\text{mod}\,\mathscr{T}),
\end{equation}
where $\vec\Omega_{\beta}=\int_{\mathfrak{p}(\beta)}^{\infty}\vec{\omega}$, and
\begin{equation}\label{eq:2.59}
\vec{\phi}(m)=\sum_{j=1}^{g}\int_{\mathfrak{p}_0}^{\mathfrak{p}(\nu_j(m))}\vec{\omega}=\mathscr{A}\big(\Sigma_{j=1}^g\mathfrak{p}(\nu_j(m))\big),
\end{equation}
which is the Abel-Jacobi variable \eqref{eq:2.25} along the $S^{m}_{\beta}$-flow.

Hence, the Baker-Akhiezer function $\mathfrak{h}_{\beta}^{(2)}(m,\mathfrak{p})$ can be constructed, in terms of the Riemann theta functions corresponding to the Riemann surface determined by the spectral curve \eqref{eq:2.6} \cite{Baker,Akhiezer,Cao,Farkas,Griffiths,Mumford},
\begin{equation}\label{eq:2.60}
\mathfrak{h}_{\beta}^{(2)}(m,\mathfrak{p})=C_{m}\cdot\frac{\theta[-\mathscr{A}(\mathfrak{p})+\vec{\phi}(m)+\vec{K}]}
{\theta[-\mathscr{A}(\mathfrak{p})+\vec{\phi}(0)+\vec{K}]}e^{m\int_{\mathfrak{p}_0}^{\mathfrak{p}}\omega[\mathfrak{p}(\beta),\infty]},
\end{equation}
where $C_{m}, \vec{K}$  are constants, independent of $\mathfrak{p}\in\mathcal{R}$. By letting $\mathfrak{p} \rightarrow\infty$ in equation \eqref{eq:2.60}, with the help of Lemma 2.3, we solve the constant factor as
\begin{equation}\label{eq:2.61}
C_{m}= \displaystyle\frac{ \theta[-\mathscr{A}(\infty) + \vec{\phi}(0)
+\vec{ K}]}{ \theta[-\mathscr{A}(\infty)
+\vec{ \phi}(m) + \vec{K}]} \cdot \displaystyle\frac{1}{(r_{\beta}^{\infty})^{m}},\quad
r_{\beta}^{\infty}=\underset{ \mathfrak{p} \rightarrow\infty
}{\mathrm{lim}}
t(\mathfrak{p})e^{\int_{\mathfrak{p}_0}^{\mathfrak{p}}\omega[\mathfrak{p}(\beta),\infty]}.
\end{equation}
Thus,
\begin{equation}\label{eq:2.62}
\mathfrak{h}_{\beta}^{(2)}(m,\mathfrak{p})= \displaystyle\frac{
\theta[-\mathscr{A}(\mathfrak{p}) + \vec{\phi}(m) + \vec{K}]}{ \theta[-\mathscr{A}(\infty) + \vec{\phi}(m) + \vec{K}]} \cdot \displaystyle\frac{
\theta[-\mathscr{A}(\infty) + \vec{\phi}(0) + \vec{K}]}{ \theta[-\mathscr{A}(\mathfrak{p}) + \vec{\phi}(0) + \vec{K}]}\cdot \Big(\displaystyle\frac{1}{r_{\beta}^{\infty}}
e^{\int_{\mathfrak{p}_0}^{\mathfrak{p}}\omega[\mathfrak{p}(\beta),\infty]}\Big)^{m}.
\end{equation}

In \cite{Baker,Akhiezer,Krichever,Belokolos,Matveev,Gesztesy}, it indicates that the reconstruction of some finite-gap potentials can be reduced to the
classical Jacobi inversion problem on hyperelliptic Riemann surfaces. In the present case, we reconstruct the discrete potentials $a_{m}, b_{m}$ by using the expression \eqref{eq:2.62}, even through the constraint \eqref{eq:2.35} for them has been given. Then the finite genus solutions to the lpKdV equation \eqref{eq:1.1} can be deduced \cite{Cao}. Hence, we consider equation \eqref{eq:2.45} implying
\begin{eqnarray*}
&&\mathfrak{h}_{\beta}^{(1)}(m+1,\mathfrak{p})=a_{m} \mathfrak{h}_{\beta}^{(1)}(m,\mathfrak{p})+(-\lambda+\beta+a_{m}b_{m})\mathfrak{h}_{\beta}^{(2)}(m,\mathfrak{p}),\\
&&\mathfrak{h}_{\beta}^{(2)}(m+1,\mathfrak{p})= \mathfrak{h}_{\beta}^{(1)}(m,\mathfrak{p})+b_{m}\mathfrak{h}_{\beta}^{(2)}(m,\mathfrak{p}).
\end{eqnarray*}
After eliminating $\mathfrak{h}_{\beta}^{(1)}(m,\mathfrak{p})$, we have
\begin{equation}\label{eq:2.63}
\mathfrak{h}_{\beta}^{(2)}(m+1,\mathfrak{p})= (b_{m}+a_{m-1})\mathfrak{h}_{\beta}^{(2)}(m,\mathfrak{p})-(\lambda-\beta)\mathfrak{h}_{\beta}^{(2)}(m-1,\mathfrak{p}).
\end{equation}
Note that the relation (\ref{eq:2.40}) between $a_{m}$ and $b_{m}$ is not enough for further calculations. Actually, we need to combine it with the compatibility of spectral problems (\ref{eq:1.2}) and (\ref{eq:1.3}). Then the following relations are obtained:
\begin{eqnarray}\label{eq:2.64}
&&a_{m}=z_{m}+v_{m+1}, \\\label{eq:2.65}
&&b_{m}=z_{m}-v_{m},\\\label{eq:2.66}
&&(z_{m}+z_{m-1})_{x}=z_{m}^{2}-z_{m-1}^{2},
\end{eqnarray}
where $z_{m}=\sqrt{v_{m+1}^{2}+v_{m}^{2}-\beta}$.

Remark: Equation (\ref{eq:2.66}) can also be derived from a B\"acklund transformation for the potential KdV equation \cite{Hietarinta,Wahlquist},
\begin{equation*}
(\mathfrak{u}_{m+1}+\mathfrak{u}_{m})_{x}=2\lambda-\displaystyle\frac{1}{2}(\mathfrak{u}_{m+1}-\mathfrak{u}_{m})^{2},
\end{equation*}
when selecting
\begin{equation}\label{eq:2.67}
-2z_{m}=\mathfrak{u}_{m+1}-\mathfrak{u}_{m}.
\end{equation}
According to the permutability property of the B\"acklund transformations, $\mathfrak{u}$ satisfies the lpKdV equation.

Hence, we are supposed to integrate $z_{m}$. By equations (\ref{eq:2.64}) and (\ref{eq:2.65}), the coefficient $b_{m}+a_{m-1}$ in equation (\ref{eq:2.63}) can be written as $z_{m}+z_{m-1}$. Now we calculate $z_{m}+z_{m-1}$ in two ways. First, we have
\begin{equation}\label{eq:2.68}
z_{m}+z_{m-1}= \underset{\mathfrak{p} \rightarrow
\infty} {\mathrm{lim}}\Big(\displaystyle\frac{
\mathfrak{h}_{\beta}^{(2)}(m+1,\mathfrak{p})}{
\mathfrak{h}_{\beta}^{(2)}(m,\mathfrak{p})} + \displaystyle\frac{
\lambda(\mathfrak{p})\mathfrak{h}_{\beta}^{(2)}(m-1,\mathfrak{p})}{
\mathfrak{h}_{\beta}^{(2)}(m,\mathfrak{p})}\Big),
\end{equation}
and with the help of equation \eqref{eq:2.15}, we get
\begin{equation}\label{eq:2.69}
\displaystyle\frac{ \theta[-\mathscr{A}(\mathfrak{p}) + \vec{\phi}(m) +
\vec{K}]}{\theta[-\mathscr{A}(\infty) +\vec{\phi}(m) +
\vec{K}]}=1-t\Theta_{m}+O(t^{2}),
\end{equation}
where~$\Theta_{m}=\partial_{x}\mid_{x=0}\log\theta[x\vec{\Omega}_{1}+\vec{K}(m)]$,
$\vec{K}(m)=\eta+\vec{\phi}(m)+\vec{K}$ with $\eta$ given by \eqref{eq:2.18}. Thus,
\begin{eqnarray*}
&&\displaystyle\frac{
\mathfrak{h}_{\beta}^{(2)}(m+1,\mathfrak{p})}{
\mathfrak{h}_{\beta}^{(2)}(m,\mathfrak{p})}=\displaystyle\frac{1}{t}\{1+[\Theta_{m}-\Theta_{m+1}+\epsilon_{\beta}]t+O(t^{2})\},\\
 &&\displaystyle\frac{
\lambda(\mathfrak{p})\mathfrak{h}_{\beta}^{(2)}(m-1,\mathfrak{p})}{
\mathfrak{h}_{\beta}^{(2)}(m,\mathfrak{p})}=\displaystyle\frac{1}{t}\{-1+[\Theta_{m-1}-\Theta_{m}+\epsilon_{\beta}]t+O(t^{2})\},
\end{eqnarray*}
where $\epsilon_{\beta}$ is given by
$$
\displaystyle\frac{t}{r_{\beta}^{\infty}}
e^{\int_{\mathfrak{p}_{0}}^{\mathfrak{p}}\omega[\mathfrak{p}(\beta),\infty]}
=1+\epsilon_{\beta}t+O(t^{2}).$$
Therefore, we have
\begin{equation}\label{eq:2.70}
z_{m}+z_{m-1}=\Theta_{m-1}-\Theta_{m+1}+2\epsilon_{\beta}.
\end{equation}
Second, we have
\begin{equation}\label{eq:2.71}
\begin{split}
z_{m}+z_{m-1}&= \underset{\mathfrak{p}
\rightarrow\mathfrak{p}(\beta) }
{\mathrm{lim}}\displaystyle\frac{(\lambda-\beta)
\mathfrak{h}_{\beta}^{(2)}(m-1,\mathfrak{p})}{
\mathfrak{h}_{\beta}^{(2)}(m,\mathfrak{p})}\\
&=\displaystyle\frac{r_{\beta}^{\infty}}{r_{\beta}}\cdot\displaystyle\frac{\theta^{2}[\vec{K}(m)]}{\theta[\vec{K}(m+1)]\theta[\vec{K}(m-1)]},
\end{split}
\end{equation}
with
\begin{equation*}
r_{\beta}=\underset{\mathfrak{p}
\rightarrow\mathfrak{p}(\beta) }
{\mathrm{lim}}\displaystyle\frac{1}{\lambda-\beta}e^{\int_{\mathfrak{p}_{0}}^{\mathfrak{p}}
\omega[\mathfrak{p}(\beta),\infty]}.
 \end{equation*}
then by equation \eqref{eq:2.66}, we obtain
\begin{equation}\label{eq:2.72}
z_{m}-z_{m-1}=2\Theta_{m}-\Theta_{m+1}-\Theta_{m-1}.
\end{equation}
As a result, by adding \eqref{eq:2.70} and \eqref{eq:2.72}, we arrive at the explicit formula
\begin{equation}\label{eq:2.73}
\begin{split}
z_{m}&=\Theta_{m}-\Theta_{m+1}+\epsilon_{\beta}\\
&=\partial_{x}\mid_{x=0}\log\displaystyle\frac{\theta[x\vec{\Omega}_{1}+\vec{K}(m)]}{\theta[x\vec{\Omega}_{1}+\vec{K}(m+1)]}+\epsilon_{\beta}.
\end{split}
 \end{equation}
Here the exact expressions for $a_{m}$ and $b_{m}$ are not obtained. However, the formula \eqref{eq:2.73} is enough when we compute the solutions for the lpKdV equation \eqref{eq:1.1} in terms of Riemann theta functions.

Let the parameters $\beta=\beta_1,\,\beta_2$ be distinct and non-zero, and applying the theory in Section 2.2 to the
two parameter cases respectively, the resulting integrable maps $S_{\beta_1},\,S_{\beta_2}$ possess the same Liouville
set of integrals $F_1,\cdots,F_N$ which subsequently determine the action-angle variables $(I,\varphi)$ with $I=I(F_1,\cdots,F_N)$. Thus, in the neighborhood of each level set
\begin{equation*}
\mathcal{M}_{c}=\{(p,q)\in \mathbb{R}^{2N}: F_{1}(p,q)=c_{1},\ldots,F_{N}(p,q)=c_{N}\},
\end{equation*}
the phase flows $S_{\beta_{1}}^{m}$ and $S_{\beta_{2}}^{n}$ are linearised by the same action-angle variables \cite{Cao5}. As a corollary of the discrete version of the Liouville-Arnold theorem \cite{Suris,Veselov,Bruschi}, $S_{\beta_{1}}^{m}$ and $S_{\beta_{2}}^{n}$ commute. Then we get a well defined function,
and it can be expressed in two ways, respectively, as
\begin{align}\label{eq:2.74}
\begin{split}
\big(p(m,n),q(m,n)\big)&=S_{\beta_{1}}^{m}S_{\beta_{2}}^{n}(p_{0},q_{0})=S_{\beta_{1}}^{m}\big(p(0,n),q(0,n)\big)\\
&=S_{\beta_{2}}^{n}S_{\beta_{1}}^{m}(p_{0},q_{0})=S_{\beta_{2}}^{n}\big(p(m,0),q(m,0)\big).
\end{split}
\end{align}
Thus by equations \eqref{eq:2.29} in the two special cases, \eqref{eq:2.64} and \eqref{eq:2.65}, the $j$-th
component satisfies two equations simultaneously with
$\lambda=\alpha_j$,
\begin{align}\label{eq:2.75}
\begin{split}
&\begin{pmatrix}\tilde{p}_{j}\\\tilde{q}_{j}\end{pmatrix}
=(\alpha_{j}-\beta_{1})^{-1/2}D^{(\beta_{1})}(\alpha_{j};z^{\prime}+\tilde{v},z^{\prime}-v)\begin{pmatrix}p_{j}\\ q_{j}\end{pmatrix},\ \ z^{\prime}=\sqrt{\tilde{v}^{2}+v^{2}-\beta_{1}} ,\\
&\begin{pmatrix}\bar{p}_{j}\\ \bar{q}_{j}\end{pmatrix}
=(\alpha_{j}-\beta_{2})^{-1/2}D^{(\beta_{2})}(\alpha_{j};z^{\prime\prime}+\bar{v},z^{\prime\prime}-v)\begin{pmatrix}p_{j}\\
q_{j}\end{pmatrix},\ \ z^{\prime\prime}=\sqrt{\bar{v}^{2}+v^{2}-\beta_{2}} .
\end{split}
\end{align}
Besides the evolution of equation \eqref{eq:2.58} along the above two discrete flows gives
\begin{equation*}
\vec{\phi}(m,n)=\vec{\phi}(0,0)+m\vec\Omega_{\beta_{1}}+n\vec\Omega_{\beta_{2}}.
\end{equation*}
Comparing equation \eqref{eq:2.67} and the theta function expression \eqref{eq:2.73} of $z_{m}$, we now define
\begin{equation}\label{eq:2.76}
Z_{mn}=\partial_{x}\mid_{x=0}\log\theta(x\vec{\Omega}_{1}+
m\vec{\Omega}_{\beta_{1}}+n\vec{\Omega}_{\beta_{2}}+\vec{K}_{00}),
\end{equation}
with $\vec{K}_{00}=\eta+\vec{\phi}(0,0)+\vec{K}$. Then we have
\begin{align*}
&z^{\prime}=Z_{mn}-\tilde{Z}_{mn}+\epsilon_{\beta_{1}},\\
&z^{\prime\prime}=Z_{mn}-\bar{Z}_{mn}+\epsilon_{\beta_{2}},
\end{align*}
and straightforward calculations tell us that $(\bar{z^{\prime}})^{2}-(\tilde{z^{\prime\prime}})^{2}=(z^{\prime\prime})^{2}-(z^{\prime})^{2}-2(\beta_{1}-\beta_{2})$.
The latter relations can be used to calculate the commutator
\begin{equation}\label{eq:2.77}
\bar{D}^{(\beta_{1})}D^{(\beta_{2})}-\tilde{D}^{(\beta_{2})}D^{(\beta_{1})}=
\begin{pmatrix}1&-\bar{\tilde{Z}}_{mn}+\tilde{Z}_{mn}+\bar{Z}_{mn}-Z_{mn}+\bar{\tilde{v}}+v\\
0&1\end{pmatrix}\Xi,
\end{equation}
where
\begin{equation*}
\Xi=(\tilde{Z}_{mn}-\bar{Z}_{mn}+\epsilon_{\beta_{2}}-\epsilon_{\beta_{1}})
(Z_{mn}-\bar{\tilde{Z}}_{mn}+\epsilon_{\beta_{2}}+\epsilon_{\beta_{1}})+\beta_{2}-\beta_{1}.
\end{equation*}
\textbf{Proposition 2.4.} The lpKdV equation \eqref{eq:1.1} has the finite genus solutions
\begin{equation}\label{eq:2.78}
u(m,n)=\partial_{x}\mid_{x=0}\log\theta(x\vec{\Omega}_{1}+
m\vec{\Omega}_{\beta_{1}}+n\vec{\Omega}_{\beta_{2}}+\vec{K}_{00})-m\epsilon_{\beta_{1}}-n\epsilon_{\beta_{2}}.
\end{equation}
\noindent\emph{Proof.} The commutativity of the flow $S^{m}_{\beta_{1}}$ and
$S^{n}_{\beta_{2}}$ guarantees the compatibility of equation \eqref{eq:2.75}. Thus $\bar{D}^{(\beta_{1})}D^{(\beta_{2})}=\tilde{D}^{(\beta_{2})}D^{(\beta_{1})}$
which implies $\Xi=0$. This leads to lpKdV equation \eqref{eq:1.1} when choosing $u(m,n)=Z_{mn}-m\epsilon_{\beta_{1}}-n\epsilon_{\beta_{2}}$.   \hfill $\Box$

So far, from a novel Hamiltonian system different from the one in \cite{Cao}, we have succeeded in deducing the explicit analytic solutions, i.e. finite genus solutions in our case, for the lpKdV equation via integrable symplectic maps. Next we will investigate the lpmKdV equation and the lSKdV equation in a similar way.

\section{ The lattice potential modified KdV equation}\setcounter{equation}{0}

Let us now consider the lattice version of the potential mKdV equation (\ref{eq:1.4}). Note that Lax pairs for (\ref{eq:1.4}) have been written down in \cite{Capel,Alsallami},
but we have not been able to blend those linear problems with the algebro-geometric technique of nonlinearisation employed in the present paper.
Inspired by Section 2.3, here we select a different parametrization for the discrete potential $a$ given
in the Lax matrix (\ref{eq:1.6}), whereby (\ref{eq:1.4}) then arises as the compatibility condition of a pair of such linear problems associated with the shifts of the vector-function $\chi$ in the $m$ and $n$ directions, namely
\begin{equation}\label{eq:3.1}
\tilde{\chi}=(\lambda^{2}-\beta_{1}^{2})^{1/2}D^{(\beta_{1})}(\lambda)\chi,\ \
\bar{\chi}=(\lambda^{2}-\beta_{2}^{2})^{1/2}D^{(\beta_{2})}(\lambda)\chi,
\end{equation}
where $D^{(\beta_{1})}(\lambda)$ is given by
\begin{equation}\label{eq:3.2}
D^{(\beta_{1})}(\lambda)=
\begin{pmatrix}\lambda \displaystyle\frac{\tilde{u}}{u} & \beta_{1} \\
\beta_{1} & \lambda \displaystyle\frac{u}{\tilde{u}}
\end{pmatrix},
\end{equation}
and where $D^{(\beta_{2})}(\lambda)$ is given by a similar matrix obtained from (\ref{eq:3.2}) by making the replacements $\beta_{1} \rightarrow \beta_{2}$ and $\tilde{} \rightarrow \bar{}$.

In fact, we have
\begin{equation}\label{eq:3.3}
\bar{D}^{(\beta_{1})}D^{(\beta_{2})}-\tilde{D}^{(\beta_{2})}D^{(\beta_{1})}=
\begin{pmatrix} 0 & -\displaystyle \frac{\lambda}{\tilde{u}\bar{u}}\\
\displaystyle \frac{\lambda}{u\tilde{\bar{u}}}& 0\end{pmatrix}\Xi,
\end{equation}
where $\Xi=\beta_{1}(\bar{u}\tilde{\bar{u}}-u\tilde{u})-\beta_{2}(\tilde{u}\tilde{\bar{u}}-u\bar{u})$.

It turns out that almost everything that holds true for the lpKdV equation also holds true for the lpmKdV equation. We shall now discuss the integrable symplctic maps and show how to solve lpmKdV equation (\ref{eq:1.4}) via the nonlinearisation approach, which differs from previous approaches.

\subsection {An integrable Hamiltonian system}

 As the starting point for the subsequent calculations, we now review some results from \cite{Cao4}. Introducing a Lax matrix
\begin{equation}\label{eq:3.4}
L(\lambda;p,q)=
\begin{pmatrix}1/2+Q_{\lambda}(A^{2}p,q) & -\lambda
Q_{\lambda}(Ap,p) \\ \lambda Q_{\lambda}(Aq,q) &
-1/2-Q_{\lambda}(A^{2}p,q)\end{pmatrix},
\end{equation}
where $Q_{\lambda}(\xi,\eta)=<(\lambda^{2}-A^{2})^{-1}\xi,\eta>$, $A=\mathrm{diag}(\alpha_{1},\ldots,\alpha_{N})$ with $\alpha_{1}^{2},\ldots,\alpha_{N}^{2}$ distinct in pairs and non-zero. It satisfies the $r$-matrix ansatz
\begin{equation*}
\{L(\lambda)\underset{,}\otimes L(\mu)\}=[r_{12}(\lambda,\mu),L_{1}(\lambda)]-[r_{12}(\mu,\lambda),L_{2}(\mu)],
\end{equation*}
\begin{align*}
\begin{split}
r_{12}(\lambda,\mu)&=\displaystyle\frac{\lambda}{\lambda^{2}-\mu^{2}}\big(\lambda(I+\sigma_{3}\otimes\sigma_{3})+
\mu(\sigma_{1}\otimes\sigma_{1}+\sigma_{2}\otimes\sigma_{2})
\big)\\
&=\displaystyle\frac{2\lambda}{\lambda^{2}-\mu^{2}}\begin{pmatrix}\lambda
& 0&0&0 \\0&0&\mu&0\\0&\mu&0&0\\0&0&0&\lambda
\end{pmatrix},
\end{split}
\end{align*}
where $\sigma_{1},\sigma_{2},\sigma_{3}$ are the usual Pauli matrices. Considering the determinant of the Lax matrix (\ref{eq:3.4}),
\begin{equation}\label{eq:3.4a}
\mathcal{F}_{\lambda}\overset{\triangle}{=}\mathrm{det}L(\lambda;p,q)=-(1/2+Q_{\lambda}(A^{2}p,q))^{2}+\lambda^{2}Q_{\lambda}(Ap,p)Q_{\lambda}(Aq,q),
\end{equation}
we also have the evolution equation in the present case,
\begin{equation}\label{eq:3.5}
\mathrm{d}L(\mu)/\mathrm{d}t_{\lambda}=[W(\lambda,\mu),L(\mu)],\ \ W(\lambda,\mu)=\displaystyle\frac{ 2\mu}{
\lambda^{2}-\mu^{2}}\begin{pmatrix}\mu L^{11}(\lambda)&\lambda
L^{12}(\lambda)\\ \lambda L^{21}(\lambda)&-\mu
L^{11}(\lambda)\end{pmatrix},
\end{equation}
where $t_{\lambda}$ is the flow variable corresponding to the Hamiltonian function $\mathcal{F}_{\lambda}$. In a similar way as in Section 2, we obtain pairwise involutive quantities $F_{0},F_{1}\ldots, F_{N-1}$, i.e., $\{F_{j},F_{k}\}=0$, from the power series expansions
\begin{align}\label{eq:3.6}
\begin{split}
&\mathcal{F}_{\lambda}=-\displaystyle\frac{1}{4}+\sum_{j=1}^{\infty}F_{j}\lambda^{-2j},\ \ | \lambda |> \mathrm{max}\{| \alpha_{1} |,\ldots, | \alpha_{N} |\},\\
&\mathcal{F}_{\lambda}=\sum_{j=0}^{\infty}F_{-j}\lambda^{2j},\ \ | \lambda |< \mathrm{min}\{| \alpha_{1} |,\ldots, | \alpha_{N} |\},\\
\end{split}
\end{align}
where
\begin{equation*}
\begin{split}
F_{0}=&-(2<p,q>-1)^{2},\\
F_{1}=&<Ap,p><Aq,q>-<A^{2}p,q>,\\
F_{k}=&-<A^{2k}p,q>-\sum\limits_{i+j=k;i,j\geq1}<A^{2i}p,q><A^{2j}p,q>\\
&-\sum\limits_{i+j=k+1;i,j\geq1}<A^{2i-1}p,p><A^{2j-1}q,q>, \ \ (k\geq 2).
\end{split}
\end{equation*}
Besides, $\mathcal{F}_{\lambda}$ is a rational function of $\zeta=\lambda^{2}$ and can be factorized as
\begin{equation}\label{eq:3.7}
\mathcal{F}_{\lambda}
=-\displaystyle\frac{1}{4}\displaystyle\frac{R(\zeta)}{\alpha^{2}(\zeta)},
\end{equation}
where
\begin{equation*}
\alpha(\zeta)=\prod_{j=1}^{N}(\zeta-\alpha_{j}^{2}),\ \
Z(\zeta)=\prod_{k=1}^{N}(\zeta-\zeta_{k}), \ \
R(\zeta)=\alpha(\zeta)Z(\zeta),
\end{equation*}
 The relevant spectral curve is defined as
\begin{equation}\label{eq:3.8}
\mathcal {R}:\xi^{2}-R(\zeta)=0,
\end{equation}
with genus $g=N-1$ and two infinities $\infty_{+}$, $\infty_{-}$. For any $\zeta\in \mathbb{C}$, in the non-branch case (not equal to $\zeta_{j}, \alpha_{j}^{2}$) there are two corresponding points on $\mathcal {R}$:
\begin{equation*}
\mathfrak{p}(\zeta)=\big(\zeta,\xi=\sqrt{R(\zeta)}\big),\ \
(\tau\mathfrak{p})\big(\zeta)=(\zeta,\xi=-\sqrt{R(\zeta)}\big).
\end{equation*}
From the Lax matrix (\ref{eq:3.4}), we get the zeros of the off-diagonal entries, which are exactly the  elliptic variables $\mu_{j}^{2}, \nu_{j}^{2}$,
\begin{equation}\label{eq:3.9}
\begin{array}{lcl}
&&L^{12}(\lambda)=-\lambda<Ap,p>\displaystyle\frac{\mathfrak{m}(\zeta)}{\alpha(\zeta)},\
\ \mathfrak{m}(\zeta)=\Pi_{ j=1}^{g}(\zeta-\mu_{j}^{2}),\\
&&L^{21}(\lambda)=\lambda<Aq,q>\displaystyle\frac{\mathfrak{n}(\zeta)}{\alpha(\zeta)},\
\ \mathfrak{n}(\zeta)=\Pi_{ j=1}^{g}(\zeta-\nu_{j}^{2}),
\end{array}
\end{equation}
in terms of which the corresponding quasi-Abel-Jacobi variables and Abel-Jacobi variables read
\begin{align}\label{eq:3.10}
\begin{split}
&\vec\phi^\prime=\sum_{k=1}^g\int_{\mathfrak p_0}^{\mathfrak
p(\nu_k^{2})}\vec\omega^\prime,\quad\vec\phi=C\vec\phi^\prime=\mathscr
{A}(\sum_{k=1}^g\mathfrak p(\nu_k^{2})),\\
&\vec\psi^\prime=\sum_{k=1}^g\int_{\mathfrak p_0}^{\mathfrak
p(\mu_k^{2})}\vec\omega^\prime,\quad\vec\psi=C\vec\psi^\prime=\mathscr
{A}(\sum_{k=1}^g\mathfrak p(\mu_k^{2})),
\end{split}
\end{align}
where $\vec\phi^\prime=(\phi_1^\prime,\cdots,\phi_g^\prime)^T$, $\vec\psi^\prime=(\psi_1^\prime,\cdots,\psi_g^\prime)^T$, and $\vec\omega^\prime=(\omega_1^\prime,\cdots,\omega_g^\prime)^T, \ \ \omega_{l}^{\prime}=\zeta^{g-l}\mathrm{d}\zeta/2\sqrt{R(\zeta)}
\ \ (1\leq l\leq g)$.

It turns out that (\ref{eq:1.5}) can be nonlinearised to create a completely integrable Hamiltonian system possessing integrals $F_{0},F_{1}\ldots, F_{N-1}$ \cite{Cao4}. The latter is defined by the canonical equations
\begin{equation}\label{eq:3.17}
\partial_{x}\begin{pmatrix} p_{j}\\q_{j}\end {pmatrix}=
\begin{pmatrix}-\partial H_{1}/\partial q_{j}\\ \partial H_{1}/\partial
p_{j}\end {pmatrix}=\begin{pmatrix}\alpha_{j}^{2}/2
&-\alpha_{j}<Ap,p> \\ \alpha_{j}<Aq,q>
&-\alpha_{j}^{2}/2
\end{pmatrix}\begin{pmatrix}p_{j}\\q_{j}\end {pmatrix},(1\leq j\leq N),
\end{equation}
where $H_{1}=F_{1}/2$ is the first member in the expression of square root $\mathcal{H}_{\lambda}$ satisfying
\begin{equation}\label{eq:3.18}
-4\mathcal{F}_{\lambda}=(-4\mathcal{H}_{\lambda})^{2},\ \  \mathcal{H}_{\lambda}=-\frac{1}{4}+\sum_{j=1}^{\infty}H_{j}\lambda^{-2j}.
\end{equation}
The nonlinearisation procedure explained above also plays an important role in solving the (2+1)-dimensional derivative Toda equation by algebra-geometric technique \cite{Cao4}, while the nonlinearisation of the discrete spectral problem (\ref{eq:1.6}) can lead to new theta function solutions for lpmKdV equation (\ref{eq:1.4}).

\subsection {An integrable symplectic map}

Following the conclusion of Section 3.1, we now construct the integrable symplectic map arising from the $N$ copies of the discrete spectral problem (\ref{eq:1.6}),
\begin{equation}\label{eq:3.19}
\begin{pmatrix}
\tilde{p}_{j}\\
\tilde{q}_{j}
\end{pmatrix}=
(\alpha_{j}^{2}-\beta^{2})^{-1/2}D^{(\beta)}(\alpha_{j};a)
\begin{pmatrix}
 p_{j}\\
 q_{j}
\end{pmatrix}, \ \ (j=1,\ldots,N).
\end{equation}
According to the procedure of Section 2.2, we discuss the discrete Lax equation in the present case. The Lax matrix (\ref{eq:3.4}) can be rewritten as
\begin{equation*}
L(\lambda;p,q)=(\frac{1}{2}-<p,q>)\sigma_{3}+\frac{\lambda}{2}\sum\limits_{j=1}^{N} (\frac{\varepsilon_{j}}{\lambda-\alpha_{j}}+\frac{\delta_{j}}{\lambda+\alpha_{j}}),
\end{equation*}
where $\delta_{j}=\sigma_{3}\varepsilon_{j}\sigma_{3}$ satisfying $\tilde{\delta}_{j}D^{(\beta)}(-\alpha_{j})=D^{(\beta)}(-\alpha_{j})\delta_{j}$. Through direct calculations, we get
\begin{equation}\label{eq:3.20}
L(\lambda;\tilde{p},\tilde{q})
D^{(\beta)}(\lambda;a)-D^{(\beta)}(\lambda;a) L(\lambda;p,q)=-\beta
P^{(\beta)}(a; p,q) \mathrm{i}\sigma_{2},
\end{equation}
where
\begin{align}\label{eq:3.21}
\begin{split}
aP^{(\beta)}(a; p,q)&=a(<\tilde{p},\tilde{q}>+<p,q>-1)\\
&=a^{2}L^{12}(\beta)-2aL^{11}(\beta)-L^{21}(\beta).
\end{split}
\end{align}
Thus, the constraint on $a$ is derived by solving the quadratic equation $aP^{(\beta)}(a; p,q)=0$,
\begin{equation}\label{eq:3.22}
a=f_{\beta}(p,q)=\displaystyle\frac{-1}{\beta
Q_{\beta}(Ap,p)}\big(1/2
+Q_{\beta}(A^{2}p,q)\pm\frac{\sqrt{R(\beta^{2})}}{2\alpha(\beta^{2})}\big).
\end{equation}
Moreover, $\beta a$ gives two values of a single-valued meromorphic function on the curve $\mathcal {R}$ given by (\ref{eq:3.8}),
\begin{equation*}
\mathfrak{A}(\mathfrak{p})=\displaystyle\frac{-1}{
Q_{\beta}(Ap,p)}\big(1/2
+Q_{\beta}(A^{2}p,q)+\frac{\xi}{2\alpha(\beta^{2})}\big),
\end{equation*}
at the points $\mathfrak{p}(\beta^{2})$ and $(\tau\mathfrak{p})(\beta^{2})$, respectively. The constraint (\ref{eq:3.22}) leads to the nonlinear map
\begin{equation}\label{eq:3.23}
S_{\beta}: \begin{pmatrix} \tilde{p}\\ \tilde{q}\end{pmatrix}=(A^{2}-\beta^{2})^{-1/2}
\begin{pmatrix}a Ap+\beta q\\a^{-1}
Aq+\beta p\end{pmatrix}\Bigg|_{a=f_{\beta}(p,q)}.
\end{equation}
\textbf{Proposition 3.1.} The map $S_{\beta}$ of (\ref{eq:3.23}) is symplectic and Liouville integrable, i.e., $S_{\beta}^{*}(\mathrm{d}p\wedge \mathrm{d}q)=\mathrm{d}p\wedge \mathrm{d}q$, and $F_{0},F_{1}\ldots, F_{N-1}$ given by equation (\ref{eq:3.6}) satisfies $S_{\beta}^{*}F_{j}=F_{j}$.

\noindent\emph{Proof.} Substituting (\ref{eq:3.22}) into (\ref{eq:3.20}), we obtain
\begin{equation}\label{eq:3.24}
L(\lambda;\tilde{p},\tilde{q})
D^{(\beta)}(\lambda;a)-D^{(\beta)}(\lambda;a) L(\lambda;p,q)=0.
\end{equation}
Thus $\mathrm{det}L(\lambda;\tilde{p},\tilde{q})=\mathrm{det}L(\lambda;p,q)$ by taking the determinant, which implies $S_{\beta}^{*}F_{j}=F_{j}$.

\noindent The symplectic property is confirmed by the expression
\begin{align}\label{eq:3.25}
\begin{split}
S_{\beta}^{*}(\mathrm{d}p\wedge \mathrm{d}q)-\mathrm{d}p\wedge \mathrm{d}q
&=\sum\limits_{j=1}^{N} (\mathrm{d}\tilde{p}_{j} \wedge
\mathrm{d}\tilde{q}_{j}-\mathrm{d}p_{j} \wedge \mathrm{d}
q_{j})\\
&=\displaystyle\frac{1}{2a} \mathrm{d}a \wedge
\mathrm{d}P^{(\beta)}(a; p,q).
\end{split}
\end{align}
which is derived from equation (\ref{eq:3.19}).   \hfill $\Box$

We now define the discrete orbit $\big(p(m),q(m)\big) =
S^{m}_{\beta}(p_{0},q_{0})$. This is more discernible if we reformulate the potentials
$a(m)=a_{m}$ and $u(m)=u_{m}$ as
\begin{align}\label{eq:3.26}
\begin{split}
&a(m) = f_{\beta}\big(p(m),q(m)\big) =
(S_{\beta}^{m})^{*}f_{\beta}(p_{0},q_{0}),\\
&\tilde{u}/u = a,
 \ \ \mathrm{or}  \ \ u_{m+1}/u_{m} = a_{m}.
\end{split}
\end{align}
On the $S^{m}_{\beta}$-flow, equation (\ref{eq:3.24}) is rewritten as
\begin{equation}\label{eq:3.27}
L_{m+1}(\lambda)D^{(\beta)}_{m}(\lambda) =
D^{(\beta)}_{m}(\lambda)L_{m}(\lambda),
\end{equation}
where $L_{m}(\lambda)=L(\lambda;p(m),q(m)), D^{(\beta)}_{m}(\lambda)=D^{(\beta)}_{m}(\lambda;a_{m})$.

\noindent Then by equation (\ref{eq:3.10}), the Abel-Jacobi variables in the Jacobi variety $J(\mathcal {R})=\mathbb{C}^{g}/\mathscr T$ can be defined as
\begin{align}\label{eq:3.28}
\begin{split}
&\vec{\phi}(m) = \mathcal
{A}\big(\sum_{j=1}^{g}\mathfrak{p}(\nu_{j}^{2}(m))\big)=\sum_{j=1}^{g}\int_{\mathfrak{p}_{0}}^{\mathfrak{p}(\nu_{j}^{2}(m))}\vec{\omega},\\
&\vec{\psi}(m) = \mathcal
{A}\big(\sum_{j=1}^{g}\mathfrak{p}(\mu_{j}^{2}(m))\big)=\sum_{j=1}^{g}\int_{\mathfrak{p}_{0}}^{\mathfrak{p}(\mu_{j}^{2}(m))}\vec{\omega}.
\end{split}
\end{align}
Let us now introduce the discrete spectral problem with potential $a_{m}$
\begin{equation}\label{eq:3.29}
h_{\beta}(m+1,\lambda) =
D^{(\beta)}_{m}(\lambda)h_{\beta}(m,\lambda),
\end{equation}
and let $M_{\beta}(m,\lambda)$ be solution matrix with $M_{\beta}(0,\lambda)=I$. Obviously,
\begin{align}\label{eq:3.30}
\begin{split}
&M_{\beta}(m,\lambda) =
D^{(\beta)}_{m-1}(\lambda)D^{(\beta)}_{m-2}(\lambda)\ldots
D^{(\beta)}_{0}(\lambda),\\
&L_{m}(\lambda)M_{\beta}(m,\lambda) =
M_{\beta}(m,\lambda)L_{0}(\lambda),
\end{split}
\end{align}
where $\mathrm{det} M_{\beta}(m,\lambda) = (\lambda^{2}-\beta^{2})^{m}$. As $\lambda\rightarrow\infty$, we have the asymptotic behaviour
\begin{align}\label{eq:3.31}
 M_{\beta}(m,\lambda)=\begin{pmatrix}O(\lambda^{m}) & O(\lambda^{m-1})\\
 O(\lambda^{m-1}) & O(\lambda^{m})\end{pmatrix}.
\end{align}
Solving the eigenvalues of the linear map $L_{m}(\lambda)$,
\begin{align}\label{eq:3.32}
\begin{split}
&\rho^{\pm}_{\lambda} = \pm\rho_{_{\lambda}} =
\pm\displaystyle\sqrt{-\mathcal{F}_{\lambda}} =
\pm\sqrt{R(\zeta)}/2\alpha(\zeta),\\
&\rho_{\lambda} = 1/2 + O(\lambda^{-2}), \ \ (\lambda \rightarrow
\infty).
\end{split}
\end{align}
the associated eigenfunctions satisfy
\begin{align}\label{eq:3.33}
&h_{\beta,\pm}(m+1,\lambda) =
D^{(\beta)}_{m}(\lambda)h_{\beta,\pm}(m,\lambda),\\\label{eq:3.34}
&h_{\beta,\pm}(m,\lambda) =
\begin{pmatrix}h_{\beta,\pm}^{(1)}(m,\lambda)\\h_{\beta,\pm}^{(2)}(m,\lambda)\end{pmatrix}
 = M_{\beta}(m,\lambda)
 \begin{pmatrix}c_{\lambda}^{\pm}\\1 \end{pmatrix}.
\end{align}
We now study the common eigenvectors $h_{\beta,\pm}(m,\lambda)$ for operators $L_{m}(\lambda)$ and $D^{(\beta)}_{m}(\lambda)$ with the help of the Baker-Akhiezer functions expressed by the theta functions of hyperelliptic Riemann surface defined by the curve $\mathcal{R}$.

\noindent Since $h_{\beta,\pm}(0,\lambda) = (c^{\pm}_{\lambda}, 1)^{T}$, the relevant entries $c^{\pm}_{\lambda}$ are given by
\begin{equation}\label{eq:3.35}
c^{\pm}_{\lambda}=\displaystyle\frac{L^{11}_{0}(\lambda) \pm
\rho_{_{\lambda}}}{L^{21}_{0}(\lambda)} =
-\displaystyle\frac{L^{12}_{0}(\lambda)}{L^{11}_{0}(\lambda)\mp\rho_{_{\lambda}}},
 \ \ c^{+}_{\lambda}c^{-}_{\lambda} =-
\displaystyle\frac{L^{12}_{0}(\lambda)}{L^{21}_{0}(\lambda)},
\end{equation}
which as $\lambda \rightarrow \infty$, behave as
\begin{align}\label{eq:3.36}
\begin{split}
&c^{+}_{\lambda}=\displaystyle\frac{\lambda}{<Aq,q>\mid_{0}}(1+ O(\lambda^{-2})),\\
&c^{-}_{\lambda}=\displaystyle\frac{<Ap,p>\mid_{0}}{\lambda}(1+ O(\lambda^{-2})).
\end{split}
\end{align}
Furthermore, $\lambda c^{+}_{\lambda}$ and $\lambda c^{-}_{\lambda}$ are the values of a meromorphic function on $\mathcal{R}$,
\begin{equation*}
\mathcal{C}(\mathfrak{p})=\displaystyle\frac{-\zeta<(\zeta-A^{2})^{-1}Ap_{0},p_{0}>}{-1/2-<(\zeta-A^{2})^{-1}A^{2}p_{0},q_{0}>+\xi/2\alpha(\zeta)},
\end{equation*}
at the points $\mathfrak{p}(\lambda^{2})$ and $(\tau\mathfrak{p})(\lambda^{2})$, respectively.

Quite similarly as in Section 2, relying on equations (\ref{eq:3.9}), (\ref{eq:3.30}), (\ref{eq:3.34}) and (\ref{eq:3.35}) we have in this case the following formulas:
\begin{align}\label{eq:3.37}
\begin{split}
h^{(1)}_{\beta,+}(m,\lambda) \cdot h^{(1)}_{\beta,-}(m,\lambda)
=&\displaystyle\frac{-L^{12}_{m}(\lambda)}{L^{21}_{0}(\lambda)}(\zeta-\beta^{2})^{m}=\displaystyle\frac{<Ap,p>\mid_{m}}{<Aq,q>\mid_{0}}
(\zeta-\beta^{2})^{m}\prod\limits_{j=1}^{g}
\displaystyle\frac{\zeta-\mu_{j}^{2}(m)}{\zeta-\nu_{j}^{2}(0)},\\
h^{(2)}_{\beta,+}(m,\lambda) \cdot h^{(2)}_{\beta,-}(m,\lambda)
=&\displaystyle\frac{L^{21}_{m}(\lambda)}{L^{21}_{0}(\lambda)}(\zeta-\beta^{2})^{m}=\displaystyle\frac{<Aq,q>\mid_{m}}{<Aq,q>\mid_{0}}
(\zeta-\beta^{2})^{m}\prod\limits_{j=1}^{g}
\displaystyle\frac{\zeta-\nu_{j}^{2}(m)}{\zeta-\nu_{j}^{2}(0)}.
\end{split}
\end{align}
As $\lambda \rightarrow \infty$, from equations (\ref{eq:3.31}), (\ref{eq:3.34}) and (\ref{eq:3.36}) we find their asymptptic behaviours,
\begin{align}\label{eq:3.38}
\begin{split}
&h_{\beta,+}^{(1)}(m,\lambda)=\displaystyle\frac{u_{m}}{<Aq,q>\mid_{0}u_{0}}\lambda^{m+1}+O(\lambda^{m-1}),\\
& h_{\beta,-}^{(1)}(m,\lambda)=O(\lambda^{m-1}),\\
&h^{(2)}_{\beta,+}(m,\lambda)=O(\lambda^{m}),\\
&h^{(2)}_{\beta,-}(m,\lambda)=\displaystyle\frac{u_{0}}{u_{m}}\lambda^{m}+O(\lambda^{m-2}).
\end{split}
\end{align}
To get well-defined meromorphic functions on $\mathcal{R}$,  we separate the two cases of odd and even $m$, i.e., $m=2k-1,2k$, then put equation (\ref{eq:3.34}) in the form
\begin{align}\label{eq:3.39}
\begin{split}
&h_{\beta,\pm}^{(1)}(2k-1,\lambda)=\lambda c^{\pm}_{\lambda}[\lambda^{-1}M_{\beta}^{11}(2k-1,\lambda)]+ M_{\beta}^{12}(2k-1,\lambda),\\
& \lambda h_{\beta,\pm}^{(2)}(2k-1,\lambda)=\lambda c^{\pm}_{\lambda}M_{\beta}^{21}(2k-1,\lambda)+\lambda M_{\beta}^{22}(2k-1,\lambda),\\
&\lambda h_{\beta,\pm}^{(1)}(2k,\lambda)=\lambda c^{\pm}_{\lambda}M_{\beta}^{11}(2k,\lambda)+\lambda M_{\beta}^{12}(2k,\lambda),\\
& h_{\beta,\pm}^{(2)}(2k,\lambda)=\lambda c^{\pm}_{\lambda}[\lambda^{-1}M_{\beta}^{21}(2k,\lambda)]+ M_{\beta}^{22}(2k,\lambda).
\end{split}
\end{align}
Apart from $\lambda c^{\pm}_{\lambda}$, the remaining functions $M_{\beta}^{ij}$ appearing on the right-hand sides are polynomials of the argument $\zeta=\lambda^{2}$. Thus, four meromorphic functions on $\mathcal{R}$ can be obtained, with the values at $\mathfrak{p}$ and $\tau\mathfrak{p}$ given as
\begin{align}\label{eq:3.40}
\begin{split}
&\mathfrak{h}_{\beta}^{(1)}(2k-1,\mathfrak{p}(\lambda^{2}))=h_{\beta,+}^{(1)}(2k-1,\lambda), \ \
\mathfrak{h}_{\beta}^{(1)}(2k-1,\tau\mathfrak{p}(\lambda^{2}))=h_{\beta,-}^{(1)}(2k-1,\lambda), \\
&\mathfrak{h}_{\beta}^{(2)}(2k-1,\mathfrak{p}(\lambda^{2}))=\lambda h_{\beta,+}^{(2)}(2k-1,\lambda), \ \
\mathfrak{h}_{\beta}^{(2)}(2k-1,\tau\mathfrak{p}(\lambda^{2}))=\lambda h_{\beta,-}^{(2)}(2k-1,\lambda),\\
&\mathfrak{h}_{\beta}^{(1)}(2k,\mathfrak{p}(\lambda^{2}))=\lambda h_{\beta,+}^{(1)}(2k,\lambda), \ \
\mathfrak{h}_{\beta}^{(1)}(2k,\tau\mathfrak{p}(\lambda^{2}))=\lambda h_{\beta,-}^{(1)}(2k,\lambda), \\
&\mathfrak{h}_{\beta}^{(2)}(2k,\mathfrak{p}(\lambda^{2}))= h_{\beta,+}^{(2)}(2k,\lambda), \ \
\mathfrak{h}_{\beta}^{(2)}(2k,\tau\mathfrak{p}(\lambda^{2}))= h_{\beta,-}^{(2)}(2k,\lambda).
\end{split}
\end{align}
Then by using equation (\ref{eq:3.37}), we get
\begin{align}\label{eq:3.41}
\begin{split}
&\mathfrak{h}_{\beta}^{(1)}(2k-1,\mathfrak{p}(\lambda^{2}))\mathfrak{h}_{\beta}^{(1)}(2k-1,\tau\mathfrak{p}(\lambda^{2}))
=\displaystyle\frac{<Ap,p>\mid_{2k-1}}{<Aq,q>\mid_{0}}(\zeta-\beta^{2})^{2k-1}\prod\limits_{j=1}^{g}
\displaystyle\frac{\zeta-\mu_{j}^{2}(2k-1)}{\zeta-\nu_{j}^{2}(0)},\\
&\mathfrak{h}_{\beta}^{(2)}(2k-1,\mathfrak{p}(\lambda^{2}))\mathfrak{h}_{\beta}^{(2)}(2k-1,\tau\mathfrak{p}(\lambda^{2}))
=\displaystyle\frac{<Aq,q>\mid_{2k-1}}{<Aq,q>\mid_{0}}\zeta(\zeta-\beta^{2})^{2k-1}\prod\limits_{j=1}^{g}
\displaystyle\frac{\zeta-\nu_{j}^{2}(2k-1)}{\zeta-\nu_{j}^{2}(0)},\\
&\mathfrak{h}_{\beta}^{(1)}(2k,\mathfrak{p}(\lambda^{2}))\mathfrak{h}_{\beta}^{(1)}(2k,\tau\mathfrak{p}(\lambda^{2}))
=\displaystyle\frac{<Ap,p>\mid_{2k}}{<Aq,q>\mid_{0}}\zeta(\zeta-\beta^{2})^{2k}\prod\limits_{j=1}^{g}
\displaystyle\frac{\zeta-\mu_{j}^{2}(2k)}{\zeta-\nu_{j}^{2}(0)},\\
&\mathfrak{h}_{\beta}^{(2)}(2k,\mathfrak{p}(\lambda^{2}))\mathfrak{h}_{\beta}^{(2)}(2k,\tau\mathfrak{p}(\lambda^{2}))
=\displaystyle\frac{<Aq,q>\mid_{2k}}{<Aq,q>\mid_{0}}(\zeta-\beta^{2})^{2k}\prod\limits_{j=1}^{g}
\displaystyle\frac{\zeta-\nu_{j}^{2}(2k)}{\zeta-\nu_{j}^{2}(0)}.
\end{split}
\end{align}
Note that at the branch point $ \mathfrak{0}$, we have the local coordinate $\lambda$. Thus by using equations (\ref{eq:3.38}) and (\ref{eq:3.41}), we obtain divisors for the four meromorphic functions $\mathfrak{h}_{\beta}^{(1)}(2k-1,\mathfrak{p}), \mathfrak{h}_{\beta}^{(2)}(2k-1,\mathfrak{p}), \mathfrak{h}_{\beta}^{(1)}(2k,\mathfrak{p})$ and $\mathfrak{h}_{\beta}^{(2)}(2k,\mathfrak{p})$, respectively:
\begin{align}\label{eq:3.42}
\begin{split}
&\mathrm{Div}(\mathfrak{h}_{\beta}^{(1)}(2k-1,\mathfrak{p}))=\sum_{j=1}^{g}\big(\mathfrak{p}(\mu_{j}^{2}(2k-1))-\mathfrak{p}(\nu_{j}^{2}(0))\big)
+ (2k-1)\mathfrak{p}(\beta^{2})-k\infty_{+}-(k-1)\infty_{-},\\
&\mathrm{Div}(\mathfrak{h}_{\beta}^{(2)}(2k-1,\mathfrak{p}))=\sum_{j=1}^{g}\big(\mathfrak{p}(\nu_{j}^{2}(2k-1))-\mathfrak{p}(\nu_{j}^{2}(0))\big)
+ \{\mathfrak{p}(0)\}+(2k-1)\mathfrak{p}(\beta^{2})-k\infty_{+}-k\infty_{-},\\
&\mathrm{Div}(\mathfrak{h}_{\beta}^{(1)}(2k,\mathfrak{p}))=\sum_{j=1}^{g}\big(\mathfrak{p}(\mu_{j}^{2}(2k))-\mathfrak{p}(\nu_{j}^{2}(0))\big)
+\{\mathfrak{p}(0)\}+2k\mathfrak{p}(\beta^{2})-(k+1)\infty_{+}-k\infty_{-},\\
&\mathrm{Div}(\mathfrak{h}_{\beta}^{(2)}(2k,\mathfrak{p}))=\sum_{j=1}^{g}\big(\mathfrak{p}(\nu_{j}^{2}(2k))-\mathfrak{p}(\nu_{j}^{2}(0))\big)
+2k\mathfrak{p}(\beta^{2})-k\infty_{+}-k\infty_{-}.
\end{split}
\end{align}
Similarly as proved in Section 2.3, we put the above results in the Jacobi variety $J(\mathcal {R})$, and arrive at the evolution formula for Abel-Jacobi variables (\ref{eq:3.28}),
\begin{align}\label{eq:3.43}
\begin{split}
&\vec{\psi}(2k-1) \equiv \vec{\phi}(0) +(2k-1)\vec{\Omega}_{\beta}^{-}+k\vec{\Omega},\quad (\mathrm{mod}\mathscr T),\\
&\vec{\phi}(2k-1) \equiv \vec{\phi}(0) +(2k-1)\vec{\Omega}_{\beta}^{-}+k\vec{\Omega}+\vec{\Omega}_{0}^{-},\quad (\mathrm{mod}\mathscr T),\\
&\vec{\psi}(2k) \equiv \vec{\phi}(0) +2k\vec{\Omega}_{\beta}^{-}+(k+1)\vec{\Omega}+\vec{\Omega}_{0}^{-},\quad (\mathrm{mod}\mathscr T),\\
&\vec{\phi}(2k) \equiv \vec{\phi}(0) +2k\vec{\Omega}_{\beta}^{-}+k\vec{\Omega},\quad (\mathrm{mod}\mathscr T),
\end{split}
\end{align}
where $\vec{\Omega}_{\beta}^{-}=\int_{\mathfrak{p}(\beta^{2})}^{\infty_{-}}\vec{\omega}, \vec{\Omega}_{0}^{-}=\int_{\mathfrak{p}(0)}^{\infty_{-}}\vec{\omega},$ and $\vec{\Omega}=\int_{\infty_{-}}^{\infty_{+}}\vec{\omega}$. As a result, the theta function expressions of Baker-Akhiezer functions $\mathfrak{h}_{\beta}^{(l)}(m,\mathfrak{p})), (l = 1, 2)$, read
\begin{align}\label{eq:3.44}
\begin{split}
\mathfrak{h}_{\beta}^{(1)}(2k-1,\mathfrak{p})=&\displaystyle\frac{
\theta[-\mathscr{A}(\mathfrak{p}) + \vec{\psi}(2k-1) + \vec{K}]}{\theta[-\mathscr{A}(\mathfrak{p}) + \vec{\phi}(0) + \vec{K}]} \cdot
\displaystyle\frac{\theta[-\mathscr{A}(\infty_{+}) + \vec{\phi}(0) + \vec{K}]}{ \theta[-\mathscr{A}(\infty_{+}) + \vec{\psi}(2k-1) + \vec{K}]}\cdot \\ &\cdot\displaystyle\frac{u_{2k-1}}{<Aq,q>\mid_{0}u_{0}}\cdot\displaystyle\frac{1}{(r_{\beta}^{+})^{k}}\cdot e^{(1-k)\int_{\mathfrak{p}}^{\infty_{+}}\omega[\mathfrak{p}(\beta^{2}),\infty_{-}]+
k\int_{\mathfrak{p}_{0}}^{\mathfrak{p}}\omega[\mathfrak{p}(\beta^{2}),\infty_{+}]}, \\
\mathfrak{h}_{\beta}^{(2)}(2k-1,\mathfrak{p})=&\displaystyle\frac{
\theta[-\mathscr{A}(\mathfrak{p}) + \vec{\phi}(2k-1) + \vec{K}]}{\theta[-\mathscr{A}(\mathfrak{p}) + \vec{\phi}(0) + \vec{K}]}\cdot\displaystyle\frac{
\theta[-\mathscr{A}(\infty_{-}) + \vec{\phi}(0) + \vec{K}]}{\theta[-\mathscr{A}(\infty_{-}) + \vec{\phi}(2k-1) + \vec{K}]}\cdot \\
&\cdot\displaystyle\frac{u_{0}}{u_{2k-1}}\cdot\displaystyle\frac{1}{(r_{\beta}^{-})^{k-1}r_{0}^{-}}\cdot e^{-k\int_{\mathfrak{p}}^{\infty_{-}}\omega[\mathfrak{p}(\beta^{2}),\infty_{+}]+
\int_{\mathfrak{p}_{0}}^{\mathfrak{p}}(k-1)\omega[\mathfrak{p}(\beta^{2}),\infty_{-}]+\omega[\mathfrak{p}(0),\infty_{-}]},\\
\mathfrak{h}_{\beta}^{(1)}(2k,\mathfrak{p})=&\displaystyle\frac{
\theta[-\mathscr{A}(\mathfrak{p}) + \vec{\psi}(2k) + \vec{K}]}{\theta[-\mathscr{A}(\mathfrak{p}) + \vec{\phi}(0) + \vec{K}]} \cdot
\displaystyle\frac{\theta[-\mathscr{A}(\infty_{+}) + \vec{\phi}(0) + \vec{K}]}{ \theta[-\mathscr{A}(\infty_{+}) + \vec{\psi}(2k) + \vec{K}]}\cdot \\ &\cdot\displaystyle\frac{u_{2k}}{<Aq,q>\mid_{0}u_{0}}\cdot\displaystyle\frac{1}{(r_{\beta}^{+})^{k}r_{0}^{+}}\cdot e^{-k\int_{\mathfrak{p}}^{\infty_{+}}\omega[\mathfrak{p}(\beta^{2}),\infty_{-}]+
\int_{\mathfrak{p}_{0}}^{\mathfrak{p}}k\omega[\mathfrak{p}(\beta^{2}),\infty_{+}]+\omega[\mathfrak{p}(0),\infty_{+}]}, \\
\mathfrak{h}_{\beta}^{(2)}(2k,\mathfrak{p})=&\displaystyle\frac{
\theta[-\mathscr{A}(\mathfrak{p}) + \vec{\phi}(2k) + \vec{K}]}{\theta[-\mathscr{A}(\mathfrak{p}) + \vec{\phi}(0) + \vec{K}]} \cdot
\displaystyle\frac{\theta[-\mathscr{A}(\infty_{-}) + \vec{\phi}(0) + \vec{K}]}{ \theta[-\mathscr{A}(\infty_{-}) + \vec{\phi}(2k) + \vec{K}]}\cdot \\ &\cdot\displaystyle\frac{u_{0}}{u_{2k}}\cdot\displaystyle\frac{1}{(r_{\beta}^{-})^{k}}\cdot e^{-k\int_{\mathfrak{p}}^{\infty_{-}}\omega[\mathfrak{p}(\beta^{2}),\infty_{+}]+
k\int_{\mathfrak{p}_{0}}^{\mathfrak{p}}\omega[\mathfrak{p}(\beta^{2}),\infty_{-}]},
\end{split}
\end{align}
where
\begin{eqnarray*}
&&r_{0}^{+}=\underset{\mathfrak{p} \rightarrow \infty^{+}}
\lim\displaystyle\frac{1}{\zeta(\mathfrak{p})}
e^{\int_{\mathfrak{p}_{0}}^{\mathfrak{p}}\omega[\mathfrak{p}(0),\infty_{+}]},\
\ r_{0}^{-}=\underset{\mathfrak{p} \rightarrow \infty^{-}}
{\mathrm{lim}}\displaystyle\frac{1}{\zeta(\mathfrak{p})}
e^{\int_{\mathfrak{p}_{0}}^{\mathfrak{p}}\omega[\mathfrak{p}(0),\infty_{-}]},\\
&& r_{\beta}^{+}=\underset{\mathfrak{p} \rightarrow \infty^{+}}
{\mathrm{lim}}\displaystyle\frac{1}{\zeta(\mathfrak{p})}
e^{\int_{\mathfrak{p}_{0}}^{\mathfrak{p}}\omega[\mathfrak{p}(\beta^{2}),\infty_{+}]},\
\  r_{\beta}^{-}= \underset{\mathfrak{p} \rightarrow \infty^{-}}
{\mathrm{lim}}\displaystyle\frac{1}{\zeta(\mathfrak{p})}
e^{\int_{\mathfrak{p}_{0}}^{\mathfrak{p}}\omega[\mathfrak{p}(\beta^{2}),\infty_{-}]}.
\end{eqnarray*}

In this case, we meet a problem that the discrete potentials are in the expression (\ref{eq:3.44}). However, when we deduce the formulas for the potentials $u(m)$ and $a(m)$, the problem is solved with the help of the relation in (\ref{eq:3.26}). This relation arises from the parametrization for constructing the Lax pair of lpmKdV equation (\ref{eq:1.4}). According to Section 2.3, we now inverse the discrete potentials by using the above results.

\noindent \textbf{Proposition 3.2.} The potentials $u(m)$ and $a(m)$, defined by equation (\ref{eq:3.26}), have explicit
evolution formulas along the $S_{\beta}^{m}$-flow , respectively
\begin{align}\label{eq:3.45}
\begin{split}
u(m)=&u(\delta_{m})\cdot
\displaystyle\frac{\theta[(1-\delta_{m})\vec{\Omega}+(\delta_{m+1}-\delta_{m})\vec{\Omega}_{0}^{-}+\vec{K}(m)]\cdot\theta[\vec{K}(\delta_{m})+\vec{\Omega}]}
{\theta[(1-\delta_{m})\vec{\Omega}+(\delta_{m+1}-\delta_{m})\vec{\Omega}_{0}^{-}+\vec{K}(\delta_{m})]\cdot\theta[\vec{K}(m)+\vec{\Omega}]}\cdot\\
&\cdot e^{\frac{m-\delta_{m}}{2}[\delta_{m}R_{\beta}+(-1)^{m}R_{0\beta}]},
\end{split}
\end{align}
\begin{align}\label{eq:3.46}
\begin{split}
a(m)=&(a(0))^{(-1)^{m}}\cdot
\displaystyle\frac{\theta[(1-\delta_{m+1})\vec{\Omega}-(\delta_{m+1}-\delta_{m})\vec{\Omega}_{0}^{-}+\vec{K}(m+1)]\cdot\theta[\vec{K}(\delta_{m+1})+\vec{\Omega}]}
{\theta[(1-\delta_{m+1})\vec{\Omega}-(\delta_{m+1}-\delta_{m})\vec{\Omega}_{0}^{-}+\vec{K}(\delta_{m+1})]\cdot\theta[\vec{K}(m+1)+\vec{\Omega}]}\cdot\\
&\cdot
\displaystyle\frac{\theta[(1-\delta_{m})\vec{\Omega}+(\delta_{m+1}-\delta_{m})\vec{\Omega}_{0}^{-}+\vec{K}(\delta_{m})]\cdot\theta[\vec{K}(m)+\vec{\Omega}]}
{\theta[(1-\delta_{m})\vec{\Omega}+(\delta_{m+1}-\delta_{m})\vec{\Omega}_{0}^{-}+\vec{K}(m)]\cdot\theta[\vec{K}(\delta_{m})+\vec{\Omega}]}\cdot\\
& \cdot
e^{\frac{1}{2}[m(-1)^{m}+\delta_{m}]R_{\beta}+m(-1)^{m+1}R_{0\beta}},
\end{split}
\end{align}
where $\delta_{j}$ is equal to 0 and 1 for even and odd $j$ respectively, and
\begin{align}\label{eq:3.47}
\begin{split}
&\vec{K}(m)=\vec{\phi}(m)+\vec{K}+\int_{\infty_{+}}^{\mathfrak{p}_{0}}\vec{\omega},\
\ R_{\beta}=\mathrm{ln}\displaystyle\frac{r_{\beta}^{+}r_{\beta}^{\pm}}{r_{\beta}^{-}r_{\beta}^{\mp}},\\
&R_{0\beta}=(\displaystyle\int_{\mathfrak{p}_{0}}^{\mathfrak{p}(\beta^{2})}\omega[\mathfrak{p}(0),\infty_{+}]
+\omega[\mathfrak{p}(0),\infty_{-}])\cdot
\mathrm{ln}\displaystyle\frac{r_{\beta}^{+}}{\beta^{2}r_{\beta}^{\mp}r_{0}^{+}r_{0}^{-}},\\
&r_{\beta}^{\pm}=e^{\int_{\mathfrak{p}_{0}}^{\infty_{+}}\omega[\mathfrak{p}(\beta^{2}),\infty_{-}]},\
\
r_{\beta}^{\mp}=e^{\int_{\mathfrak{p}_{0}}^{\infty_{-}}\omega[\mathfrak{p}(\beta^{2}),\infty_{+}]}.
\end{split}
\end{align}
\noindent \emph{Proof.} By equation (\ref{eq:3.33}), we have
\begin{equation}\label{eq:3.48}
\begin{cases}
\mathfrak{h}_{\beta}^{(1)}(2k,\mathfrak{p})=\zeta
a_{2k-1}\mathfrak{h}_{\beta}^{(1)}(2k-1,\mathfrak{p})+\beta
\mathfrak{h}_{\beta}^{(2)}(2k-1,\mathfrak{p}),\\
\mathfrak{h}_{\beta}^{(2)}(2k,\mathfrak{p})=\beta
\mathfrak{h}_{\beta}^{(1)}(2k-1,\mathfrak{p})+a_{2k-1}^{-1}
\mathfrak{h}_{\beta}^{(2)}(2k-1,\mathfrak{p}),
\end{cases}
\end{equation}
and
\begin{equation}\label{eq:3.49}
\begin{cases}
\mathfrak{h}_{\beta}^{(1)}(2k+1,\mathfrak{p})=
a_{2k}\mathfrak{h}_{\beta}^{(1)}(2k,\mathfrak{p})+\beta
\mathfrak{h}_{\beta}^{(2)}(2k,\mathfrak{p}),\\
\mathfrak{h}_{\beta}^{(2)}(2k+1,\mathfrak{p})=\beta
\mathfrak{h}_{\beta}^{(1)}(2k,\mathfrak{p})+\zeta a_{2k}^{-1}
\mathfrak{h}_{\beta}^{(2)}(2k,\mathfrak{p}),
\end{cases}
\end{equation}
where $\mathfrak{p}=\mathfrak{p}(\zeta), \zeta=\lambda^{2}$. According to (\ref{eq:3.42}), the order of the zero $\mathfrak{p}(\beta^{2})$ of $\mathfrak{h}_{\beta}^{(l)}(m,\mathfrak{p})), (l = 1, 2)$, is equal to $m$. Thus, from the above equations we get
\begin{eqnarray*}
a_{2k-1}=
\underset{\lambda \rightarrow \beta}
{\mathrm{lim}}\displaystyle\frac{-
\mathfrak{h}_{\beta}^{(2)}(2k-1,\mathfrak{p}(\lambda^{2})) }{\beta
\mathfrak{h}_{\beta}^{(1)}(2k-1,\mathfrak{p}(\lambda^{2}))},\ \
a_{2k}= \underset{\lambda \rightarrow \beta}
{\mathrm{lim}}\displaystyle\frac{-\beta
\mathfrak{h}_{\beta}^{(2)}(2k,\mathfrak{p}(\lambda^{2})) }{
\mathfrak{h}_{\beta}^{(1)}(2k,\mathfrak{p}(\lambda^{2}))}.
\end{eqnarray*}
By using equations (\ref{eq:3.43}) and (\ref{eq:3.44}), we obtain the following relation in terms of theta functions between $u_{m}$ and $a_{m}$
\begin{align}\label{eq:3.50}
\begin{split}
a_{2k-1}=&\displaystyle\frac{
\theta[2k\vec{\Omega}_{\beta}^{-}+(k+1)\vec{\Omega} + \vec{\Omega}_{0}^{-}+\vec{K}(0)]}{\theta[2k\vec{\Omega}_{\beta}^{-}+(k+1)\vec{\Omega} + \vec{K}(0)]} \cdot
\displaystyle\frac{\theta[\vec{\Omega} + \vec{K}(0)]}{ \theta[(2k-1)\vec{\Omega}_{\beta}^{-}+(k+1)\vec{\Omega} + \vec{\Omega}_{0}^{-}+\vec{K}(0)]}\cdot \\ &\cdot\displaystyle\frac{\theta[(2k-1)\vec{\Omega}_{\beta}^{-}+k\vec{\Omega} +\vec{K}(0)]}{\theta[\vec{K}(0)]} \cdot\displaystyle\frac{<Aq,q>\mid_{0}u_{0}^{2}}{(-\beta)u_{2k-1}^{2}}\cdot\displaystyle\frac{(r_{\beta}^{+})^{k}(r_{\beta}^{\pm})^{k-1}}{(r_{\beta}^{-})^{k-1}
(r_{\beta}^{\mp})^{k}r_{0}^{-}}\cdot e^{\int_{\mathfrak{p}_{0}}^{\mathfrak{p}(\beta^{2})}\omega[\mathfrak{p}(0),\infty_{-}]}, \\
a_{2k}=&\displaystyle\frac{
\theta[(2k+1)\vec{\Omega}_{\beta}^{-}+(k+1)\vec{\Omega}+\vec{K}(0)]}{\theta[(2k+1)\vec{\Omega}_{\beta}^{-}+(k+2)\vec{\Omega}+ \vec{\Omega}_{0}^{-} +\vec{K}(0)]} \cdot
\displaystyle\frac{\theta[\vec{\Omega} + \vec{K}(0)]}{ \theta[2k\vec{\Omega}_{\beta}^{-}+(k+1)\vec{\Omega}+\vec{K}(0)]}\cdot \\ &\cdot\displaystyle\frac{\theta[2k\vec{\Omega}_{\beta}^{-}+(k+1)\vec{\Omega} + \vec{\Omega}_{0}^{-}+\vec{K}(0)]}{\theta[\vec{K}(0)]} \cdot\displaystyle\frac{(-\beta)<Aq,q>\mid_{0}u_{0}^{2}}{u_{2k}^{2}}\cdot\displaystyle\frac{(r_{\beta}^{+}r_{\beta}^{\pm})^{k}r_{0}^{+}}
{(r_{\beta}^{-}r_{\beta}^{\mp})^{k}}\cdot e^{-\int_{\mathfrak{p}_{0}}^{\mathfrak{p}(\beta^{2})}\omega[\mathfrak{p}(0),\infty_{+}]}.
\end{split}
\end{align}
Note that (\ref{eq:3.26}) gives another relation between them, i.e. $a_{2k-1}=u_{2k}/u_{2k-1}, a_{2k}=u_{2k+1}/u_{2k}$, which implies
\begin{equation}\label{eq:3.51}
\displaystyle\frac{a_{2k}}{a_{2k-1}}=\displaystyle\frac{u_{2k-1}u_{2k+1}}{u_{2k}^{2}},\ \
\displaystyle\frac{a_{2k+1}}{a_{2k}}=\displaystyle\frac{u_{2k}u_{2k+2}}{u_{2k+1}^{2}}.
\end{equation}
Substituting (\ref{eq:3.50}) into (\ref{eq:3.51}), we obtain the central result (\ref{eq:3.45}) for the solution by induction and some calculations. Then by using (\ref{eq:3.26}), equation (\ref{eq:3.46}) is obtained as well.   \hfill $\Box$

Now the theta function expression for the discrete potential $u(m)$ is written down. By which we will discuss the explicit solutions to the lpmKdV equation (\ref{eq:1.4}) through the commutativity of discrete flows.

\subsection{ The finite genus solutions to the lpmKdV equation}

Taking now any two distinct lattice parameters $\beta_{1}^{2},\beta_{2}^{2}$,
the integrable symplectic maps $S_{\beta_{1}}$ and $S_{\beta_{2}}$
share the same Liouville set of integrals, the confocal polynomials,
therefore, the resulting discrete phase flows, i.e., $S_{\beta_{1}}^{m}$- and $S_{\beta_{2}}^{n}$-flow commute. Thus a well-defined
function $\big(p(m,n),q(m,n)\big)$ is obtained, and by equation (\ref{eq:3.26}) the $j$-th component $(p_{j}(m,n),q_{j}(m,n))$ solves two copies of equation (\ref{eq:3.19}) with $\beta = \beta_{1},\beta_{2}$ simultaneously in the case of $\lambda=\alpha_{j}$,
\begin{align}\label{eq:3.52}
&\begin{pmatrix}\tilde{p}_{j}\\\tilde{q}_{j}\end{pmatrix}
=(\alpha_{j}^{2}-\beta_{1}^{2})^{-1/2}D^{(\beta_{1})}(\alpha_{j};\tilde{u}/u)\begin{pmatrix}p_{j}\\ q_{j}\end{pmatrix},\\\label{eq:3.53}
&\begin{pmatrix}\bar{p}_{j}\\ \bar{q}_{j}\end{pmatrix}
=(\alpha_{j}^{2}-\beta_{2}^{2})^{-1/2}D^{(\beta_{2})}(\alpha_{j};\bar{u}/u)\begin{pmatrix}p_{j}\\
q_{j}\end{pmatrix}.
\end{align}
The compatibility of equations (\ref{eq:3.52}) and (\ref{eq:3.53}) is provided by the commutativity of the $S_{\beta_{1}}^{m}$- and $S_{\beta_{2}}^{n}$-flow expressed by $\bar{D}^{(\beta_{1})}D^{(\beta_{2})}=\tilde{D}^{(\beta_{2})}D^{(\beta_{1})}$. Then from equation (\ref{eq:3.3}), the evolution of the function $u(m)$ given by equation (\ref{eq:3.45}) along the discrete flows yields

\noindent \textbf{Proposition 3.3.}  The lpmKdV equation (\ref{eq:1.4}) has finite genus solutions as
\begin{align}\label{eq:3.54}
\begin{split}
u(m,n)=&
u(\delta_{m},\delta_{n})\cdot\displaystyle\frac{\theta[(1-\delta_{m})\vec{\Omega}+(\delta_{m+1}-\delta_{m})\vec{\Omega}_{0}^{-}+\vec{K}(m,n)]}
{\theta[(1-\delta_{n})\vec{\Omega}+(\delta_{n+1}-\delta_{n})\vec{\Omega}_{0}^{-}+\vec{K}(\delta_{m},\delta_{n})]}\cdot\\
&\cdot\displaystyle\frac{\theta[(1-\delta_{n})\vec{\Omega}+(\delta_{n+1}-\delta_{n})\vec{\Omega}_{0}^{-}+\vec{K}(\delta_{m},n)]
\cdot\theta[\vec{K}(\delta_{m},\delta_{n})+\vec{\Omega}]}
{\theta[(1-\delta_{m})\vec{\Omega}+(\delta_{m+1}-\delta_{m})\vec{\Omega}_{0}^{-}+\vec{K}(\delta_{m},n)]\cdot\theta[\vec{K}(m,n)+\vec{\Omega}]}\cdot\\
& \cdot
e^{\frac{m-\delta_{m}}{2}[\delta_{m}R_{\beta_{1}}+(-1)^{m}R_{0\beta_{1}}]+
\frac{n-\delta_{n}}{2}[\delta_{n}R_{\beta_{2}}+(-1)^{n}R_{0\beta_{2}}]},
\end{split}
\end{align}
where $\vec{K}(m,n) = \vec{\phi}(m,n) + \vec{K} +
\int_{\infty_{+}}^{\mathfrak{p}_{0}}\vec{\omega}, \ \ \vec{\phi}(m,n) =\vec{\phi}(0,0)+
m\vec{\Omega}_{\beta_{1}}^{-}+n\vec{\Omega}_{\beta_{2}}^{-}+\displaystyle\frac{m+n+\delta_{m}+\delta_{n}}{2}\vec{\Omega}
+(\delta_{m}+\delta_{n})\vec{\Omega}_{0}^{-}$, and $R_{\beta_{k}}, R_{0\beta_{k}}$ are given by equation (\ref{eq:3.47}) with $\beta =\beta_{k}, k = 1,2$.

Up to now, the lpmKdV equation has been resolved via new integrable symplectic maps generated by a finite dimensional integrable Hamiltonian system associated with the Kaup-Newell problem, which is different from the results in \cite{Cao3}.

\section{ The lattice Schwarzian KdV equation}\setcounter{equation}{0}

Let us now study the lSKdV equation. The associated continuous spectral problem (\ref{eq:1.8}) and discrete spectral problem
(\ref{eq:1.9}) carries two potentials respectively, similar to the case of the lpKdV equation, while the way to deal with
them are somewhat different from the other cases. Nonetheless, from the earlier sections, it is evident that the discrete Lax equation plays an essential role, and that is also true in the present case. We note in passing that the lSKdV
(\ref{eq:1.7}) first appeared in \cite{Capel} as a special parameter limit of a slightly more general equation, the NQC
equation derived in \cite{NQC} in the context of the direct linearisation method. The latter quadrlateral lattice equation,
that is equivalent to the $({\rm Q3})_{\delta=0}$ equation of the ABS list \cite{Adler} found more recently, also gives rise
discrete version of the Volterra-Kac-van Moerbeke equation in special parameter and continuum limits. In the present context,
following the computations of the previous sections, we are naturally concerned with the Lax matrix, for which we take the one
of \cite{Cao6} (up to a factor $-2\lambda$), which
corresponds to the finite-dimensional Hamiltonian systems for the Kac-van Moerbeke hierarchy,
\begin{equation}\label{eq:4.1}
 L(\lambda;p,q)=
\begin{pmatrix}\lambda/2+\lambda Q_{\lambda}(p,q) & -<p,q>-
Q_{\lambda}(Ap,p) \\ 1+ Q_{\lambda}( Aq,q) &
-\lambda/2-\lambda Q_{\lambda}(p,q)\end{pmatrix},
\end{equation}
where $Q_{\lambda}(\xi,\eta)=<(\lambda^{2}-A^{2})^{-1}\xi,\eta>$, $A=\mathrm{diag}(\alpha_{1},\ldots,\alpha_{N})$ with $\alpha_{1}^{2},\ldots,\alpha_{N}^{2}$ pairwise distinct and non-zero. Moreover, we have the following linear map from (\ref{eq:1.9}):
\begin{equation}\label{eq:4.2}
\begin{pmatrix}
\tilde{p}_{j}\\
\tilde{q}_{j}
\end{pmatrix}=
(\alpha_{j}^{2}-\beta^{2})^{-1/2}D^{(\beta)}(\alpha_{j};a,s)
\begin{pmatrix}
 p_{j}\\
 q_{j}
\end{pmatrix}, \ \ (j=1,\ldots,N).
\end{equation}
Now we compute
\begin{equation}\label{eq:4.3}
\Upsilon\overset{\triangle}{=}L(\lambda;\tilde{p},\tilde{q})D^{(\beta)}(\lambda;a,s)-D^{(\beta)}(\lambda;a,s)L(\lambda;p,q),
\end{equation}
with the entries
\begin{align*}
\begin{split}
&\Upsilon^{11}=-\beta s^{-1}<\tilde{p},\tilde{q}>-\beta s+a(<\tilde{p},\tilde{q}>-<p,q>),\\
&\Upsilon^{12}=\lambda \beta s-\lambda a^{-1}<\tilde{p},\tilde{q}>+\lambda a<p,q>,\\
&\Upsilon^{21}=\lambda a-\lambda a^{-1}-\lambda\beta s^{-1},\\
&\Upsilon^{22}=\beta s^{-1}<p,q>+\beta s-a^{-1}(<\tilde{p},\tilde{q}>-<p,q>).
\end{split}
\end{align*}
By using $\Upsilon^{21}$, we choose the formula
\begin{equation}\label{eq:4.4}
s=\beta/(a-a^{-1}).
\end{equation}
On one hand, \eqref{eq:4.4} guarantees the realization of an associated integrable symplectic map; On the other hand, imposing \eqref{eq:4.4}, the spectral problem (\ref{eq:1.9}) can be written in the form
\begin{equation}\label{eq:4.5}
\tilde{\chi}=(\lambda^{2}-\beta^{2})^{-1/2}D^{(\beta)}(\lambda;a)\chi,
\ \ D^{(\beta)}(\lambda;a)=
\begin{pmatrix} \lambda a & \displaystyle\frac{\beta^{2}}{a-a^{-1}}\\
a-a^{-1} & \lambda a^{-1}
\end{pmatrix}.
\end{equation}
Now the number of the discrete potentials is reduced to one, same as the lpmKdV situation. Inspired by the construction of the Lax pair for lpmKdV equation (\ref{eq:1.4}) (see Section 3), we find the Lax pair for lSKdV equation (\ref{eq:1.7}),
\begin{equation}\label{eq:4.6}
\tilde{\chi}=D^{(\beta_{1})}(\lambda;\tilde{z}/z,\beta_{1}z\tilde{z}/(\tilde{z}^{2}-z^{2}))\chi,\ \
\bar{\chi}=D^{(\beta_{2})}(\lambda;\bar{z}/z,\beta_{2}z\bar{z}/(\bar{z}^{2}-z^{2}))\chi.
\end{equation}
Indeed, by direct calculation, we get
\begin{equation}\label{eq:4.7}
\bar{D}^{(\beta_{1})}D^{(\beta_{2})}-\tilde{D}^{(\beta_{2})}D^{(\beta_{1})}=
\begin{pmatrix} \displaystyle\frac{\tilde{\bar{z}}}{z(\tilde{\bar{z}}^{2}-\bar{z}^{2})(\tilde{\bar{z}}^{2}-\tilde{z}^{2})} &
\displaystyle \frac{\lambda z\tilde{\bar{z}}(\tilde{z}^{2}+\bar{z}^{2}-\tilde{\bar{z}}^{2}-z^{2})}{(\tilde{z}^{2}-z^{2})(\bar{z}^{2}-z^{2})
(\tilde{\bar{z}}^{2}-\tilde{z}^{2})(\tilde{\bar{z}}^{2}-\bar{z}^{2})}\\
0&  \displaystyle\frac{-z}{\tilde{\bar{z}}(\tilde{z}^{2}-z^{2})(\bar{z}^{2}-z^{2})}\end{pmatrix}\Xi,
\end{equation}
where $\Xi=\beta_{1}^{2}(\tilde{\bar{z}}^{2}-\tilde{z}^{2})(\bar{z}^{2}-z^{2})-\beta_{2}^{2}(\tilde{\bar{z}}^{2}-\bar{z}^{2})(\tilde{z}^{2}-z^{2})$, and hence the discrete zero curvature equation $\bar{D}^{(\beta_{1})}D^{(\beta_{2})}-\tilde{D}^{(\beta_{2})}D^{(\beta_{1})}=0$ implies $\Xi=0$. The lSKdV equation (\ref{eq:1.7}) can be deduced by letting $u=z^{2}$.

Following the procedure applied in the preceding sections, we now treat the lSKdV equation in a similar way.

\subsection {The integrable Hamiltonian system}

Based on the Lax matrix (\ref{eq:4.1}), we now exhibit an integrable Hamiltonian system for further calculations. The following fundamental
Poisson bracket relation links (\ref{eq:4.1}) to a classical $r$-matrix stucture,
\begin{align}\label{eq:4.8}
\begin{split}
&\{ L(\lambda)\underset{,}\otimes L(\mu)\}=[ r(\lambda,\mu), L_{1}(\lambda)]+[ r^{\prime}(\lambda,\mu), L_{2}(\mu)],\\
& r=\displaystyle\frac{2}{\lambda^{2}-\mu^{2}} P_{\mu\lambda}+ \sigma_{3}\otimes\sigma_{+}, \ \
 r^{\prime}=\displaystyle\frac{2}{\lambda^{2}-\mu^{2}} P_{\lambda\mu}- \sigma_{3}\otimes \sigma_{+},\\
& P_{\lambda\mu}=\begin{pmatrix}\lambda
& 0&0&0 \\0&0&\mu&0\\0&\mu&0&0\\0&0&0&\lambda
\end{pmatrix}.
\end{split}
\end{align}
The associated generating function reads:
\begin{equation*}
\mathcal{F}_{\lambda}=-\lambda^{2}(1/4+Q_{\lambda}^{2}(p,q)+Q_{\lambda}(p,q))+<p,q>(1+Q_{\lambda}(Aq,q))
+Q_{\lambda}(Ap,p)(1+Q_{\lambda}(Aq,q)),
\end{equation*}
and the corresponding evolution along the $t_{\lambda}$-flow for the Lax matrices reads:
\begin{equation}\label{eq:4.9}
\mathrm{d}L(\mu)/\mathrm{d}t_{\lambda}=[W(\lambda,\mu),L(\mu)],\ \ W(\lambda,\mu)=\displaystyle\frac{2\mu}{
\lambda^{2}-\mu^{2}}L(\lambda)+\big(\displaystyle\frac{2L^{11}(\lambda)}{\lambda+\mu}- L^{21}(\lambda)\big)\sigma_{3}.
\end{equation}
Considering now the power series expression,
\begin{equation}\label{eq:4.10}
\mathcal{F}_{\lambda}=-\frac{\zeta}{4}+\sum_{j=1}^{\infty}F_{j}\zeta^{-j},\ \ \zeta=\lambda^{2},
\end{equation}
 it yields two types of objects:

\noindent a) $N$ smooth functions $\{F_{j}(p, q), 1\leq j \leq N\}$ involutive with each other,
\begin{equation*}
\begin{split}
F_{1}=&-<p,q>^{2}-<A^{2}p,q>+<Ap,p>+<p,q><Aq,q>,\\
F_{j}=&-<A^{2j}p,q>+<A^{2j-1}p,p>+<p,q><A^{2j-1}q,q>\\
&-\sum\limits_{k+l+1=j;k,l\geq0}<A^{2k}p,q><A^{2l}p,q>+\sum\limits_{k+l+2=j;k,l\geq0}<A^{2k+1}p,p><A^{2l+1}q,q>, \ \ (j\geq 2).
\end{split}
\end{equation*}

\noindent b) square root $\mathcal{H}_{\lambda}$ satisfying
\begin{equation}\label{eq:4.11}
-\displaystyle \frac{4}{\lambda^{2}}\mathcal{F}_{\lambda}=(1+4\mathcal{H}_{\lambda})^{2},\ \ \mathcal{H}_{\lambda}=\sum_{j=1}^{\infty}H_{j}\zeta^{-j-1},
\end{equation}
where $H_{1}=-\displaystyle\frac{1}{2}F_{1}$, whose corresponding Hamiltonian system is
\begin{align}\label{eq:4.12}
\begin{split}
\partial_{x}\begin{pmatrix} p_{j}\\q_{j}\end {pmatrix}&=
\begin{pmatrix}-\partial H_{1}/\partial q_{j}\\ \partial H_{1}/\partial
p_{j}\end {pmatrix}\\
&=\begin{pmatrix}-\alpha_{j}^{2}/2+<p,q>+\displaystyle\frac{<Aq,q>}{2}
&\alpha_{j}<p,q> \\-\alpha_{j}
&\alpha_{j}^{2}/2-<p,q>-\displaystyle\frac{<Aq,q>}{2}
\end{pmatrix}\begin{pmatrix}p_{j}\\q_{j}\end {pmatrix},
\end{split}
\end{align}
$(1\leq j\leq N)$. Comparing with equation (\ref{eq:1.8}), we select the constraint
\begin{equation}\label{eq:4.13}
(v,w)=(<p,q>,<Aq,q>/2).
\end{equation}
In this sense, (\ref{eq:4.12}) is the nonlinearisation of (\ref{eq:1.8}).

\noindent Consider the fractional expression
\begin{align}\label{eq:4.14}
\mathcal{F}_{\lambda}=-\frac{1}{4}\frac{R(\zeta)}{\zeta\alpha^{2}(\zeta)},\ \ R(\zeta)=\zeta\alpha(\zeta)\prod_{j=1}^{N+1}(\zeta-\zeta_{j}),\ \ \alpha(\zeta)=\prod_{j=1}^{N}(\zeta-\alpha_{j}^{2}),
\end{align}
then a curve $\mathcal{R}:\xi^2=R(\zeta)$, with genus $g=N$, is obtained. The curve $\mathcal{R}$ has two infinities $\infty_{+}$, $\infty_{-}$, and branch points $\zeta_{j}, \alpha_{j}^{2},\mathfrak{0}$. And the general points on $\mathcal {R}$ are
\begin{equation*}
\mathfrak{p}(\zeta)=\big(\zeta,\xi=\sqrt{R(\zeta)}\big),\ \
(\tau\mathfrak{p})\big(\zeta)=(\zeta,\xi=-\sqrt{R(\zeta)}\big),\ \ \zeta\in \mathbb{C}.
\end{equation*}
Introducing the corresponding elliptic coordinates $\mu_{j}^{2}, \nu_{j}^{2}$:
\begin{equation}\label{eq:4.15}
\begin{array}{lcl}
&&L^{12}(\lambda)=-<p,q>\displaystyle\frac{\mathfrak{m}(\zeta)}{\alpha(\zeta)},\
\ \mathfrak{m}(\zeta)=\Pi_{ j=1}^{N}(\zeta-\mu_{j}^{2}),\\
&&L^{21}(\lambda)=\displaystyle\frac{\mathfrak{n}(\zeta)}{\alpha(\zeta)},\
\ \mathfrak{n}(\zeta)=\Pi_{ j=1}^{N}(\zeta-\nu_{j}^{2}),
\end{array}
\end{equation}
the quasi-Abel-Jacobi and Abel-Jacobi variables are defined respectively as
\begin{align}\label{eq:4.16}
\begin{split}
&\vec\phi^\prime=(\phi_1^\prime,\cdots,\phi_g^\prime)^T=\sum_{k=1}^g\int_{\mathfrak p_0}^{\mathfrak
p(\nu_k^{2})}\vec\omega^\prime,\quad\vec\phi=C\vec\phi^\prime=\mathscr
{A}(\sum_{k=1}^g\mathfrak p(\nu_k^{2})),\\
&\vec\psi^\prime=(\psi_1^\prime,\cdots,\psi_g^\prime)^T=\sum_{k=1}^g\int_{\mathfrak p_0}^{\mathfrak
p(\mu_k^{2})}\vec\omega^\prime,\quad\vec\psi=C\vec\psi^\prime=\mathscr
{A}(\sum_{k=1}^g\mathfrak p(\mu_k^{2})),
\end{split}
\end{align}
where $\vec{\omega}^\prime=(\omega_1^\prime,\cdots,\omega_g^\prime)^T, \ \
\omega^{\prime}_j=\zeta^{g-j}\text{d}\zeta/(2\sqrt{R(\zeta)})$.

Let us consider one of the entries in (\ref{eq:4.9}), namely
\begin{equation*}
\mathrm{d}L^{12}(\mu)/\mathrm{d}t_{\lambda}=2(W^{11}(\lambda,\mu)L^{12}(\mu)-W^{12}(\lambda,\mu)L^{11}(\mu)),
\end{equation*}
and setting $\mu=\mu_{k}$, then we get the Dubrovin type equations
\begin{equation}\label{eq:4.17}
\displaystyle\frac{ 1}{2\sqrt{R(\mu_{k}^{2})}}\displaystyle\frac{
\mathrm{d}(\mu_{k}^{2})}{\mathrm{
d}t_{\lambda}}=\displaystyle\frac{1}{\alpha(\zeta)}\displaystyle\frac{
\mathfrak{m}(\zeta)}{
(\zeta-\mu_{k}^{2})\mathfrak{m}^{\prime}(\mu_{k}^{2})},\quad(1\leq
k\leq g)
\end{equation}
from which we have
\begin{equation*}
\{\psi_{l}^{\prime},\mathcal{F}_{\lambda}\}=\displaystyle\frac{
\mathrm{d}\psi_{l}^{\prime}}{
\mathrm{d}t_{\lambda}}=\displaystyle\frac{1}{\alpha(\zeta)}\zeta^{g-l},\quad(1\leq
l\leq g).
\end{equation*}
Hence,
\begin{equation}\label{eq:4.18}
\sum\limits_{j=1}^{\infty}\{\psi_{l}^{\prime},F_{j}\}\zeta^{-j}=
-\sum\limits_{j=l}^{\infty}A_{j-l}\zeta^{-j},
\end{equation}
where $A_{0}=1$, $A_{j-l}=0$ $(j< l)$. Thus, we conclude
\begin{equation*}
\big(\{\psi_{l}^{'},F_{j}\}\big)_{g\times g}=
\begin{pmatrix}1&A_{1}&A_{2}&\ldots&A_{g-1}\\ \quad
&1&A_{1}&\ldots&A_{g-2}\\ \quad & \quad &\ddots&\ddots&\vdots\\
\quad& \quad & \quad&\ddots&A_{1}\\ \quad & \quad
&\quad&\quad&1\end{pmatrix}.
\end{equation*}
which implies $F_{1},\ldots,F_{N}$ are functionally independent on the phase space $\mathcal{N}=(\mathbb{R}^{2N},\mathrm{d}p\wedge \mathrm{d}q)$.

The Liouville integrability for the Hamiltonian system (\ref{eq:4.12}) is established. We will proceed by constructing integrable symplectic maps for lSKdV equation.

\subsection {The integrable symplectic map}

With the help of the formula (\ref{eq:4.4}), the map (\ref{eq:4.2}) can be written in the form
\begin{equation}\label{eq:4.19}
\begin{pmatrix}\tilde{p}_{_{j}}\\
\tilde{q}_{_{j}}\end{pmatrix}=\frac{1}{\sqrt{\alpha_{j}^{2}-\beta^{2}}} D^{(\beta)}(\alpha_{_{j}};a)\begin{pmatrix}p_{_{j}}\\
q_{_{j}}\end{pmatrix}, \ \ (1\leq j \leq N),
\end{equation}
where $D^{(\beta)}$ is given in (\ref{eq:4.5}). Moreover, equation (\ref{eq:4.3}) becomes
\begin{equation}\label{eq:4.20}
\Upsilon\overset{\triangle}{=}L(\lambda;\tilde{p},\tilde{q})D^{(\beta)}(\lambda;a)-D^{(\beta)}(\lambda;a)L(\lambda;p,q),
\end{equation}
with the components
\begin{align*}
\begin{split}
&\Upsilon^{11}=a^{-1}<\tilde{p},\tilde{q}>-a<p,q>-\displaystyle\frac{\beta^{2}}{a-a^{-1}},\\
&\Upsilon^{12}=-\lambda \Upsilon^{11},\\
&\Upsilon^{21}=0,\\
&\Upsilon^{22}=-\Upsilon^{11}.
\end{split}
\end{align*}
Here we also use $\Upsilon$ for short. Then by equations (\ref{eq:4.1}) and (\ref{eq:4.19}), we get
\begin{equation}\label{eq:4.21}
<\tilde{p},\tilde{q}>=(a^{2}-1)L^{12}(\beta)+a^{2}<p,q>+\displaystyle\frac{\beta^{2}}{a^{2}-1}(1-L^{21}(\beta))
+2\beta(\displaystyle\frac{\beta}{2}-L^{11}(\beta)).
\end{equation}
Substituting it into $\Upsilon^{11}$, we obtain
\begin{equation}\label{eq:4.22}
\Upsilon=\displaystyle\frac{P^{(\beta)}(a; p,q)}{a^{3}-a}\begin{pmatrix}1&-\lambda\\0&-1\end{pmatrix},
\end{equation}
where
\begin{equation}\label{eq:4.23}
P^{(\beta)}(a;p,q)=(a^{2}-1)^{2}L^{12}(\beta)-2\beta (a^{2}-1)L^{11}(\beta)-\beta^{2}L^{21}(\beta),
\end{equation}
which is a quadratic polynomial with respect to $a^{2}-1$. The roots to the quadratic equation $P^{(\beta)}(a;p,q)=0$ are given by
\begin{equation}\label{eq:4.24}
a^{2}-1=\displaystyle\frac{-1}{<p,q>+
Q_{\beta}(Ap,p)}\big(\beta^{2}(1/2
+Q_{\beta}(p,q))\pm\frac{\sqrt{R(\beta^{2})}}{2\alpha(\beta^{2})}\big),
\end{equation}
which are the values of a well-defined meromorphic function on $\mathcal{R}$,
\begin{equation*}
\mathfrak {D}(\mathfrak{p})=\displaystyle\frac{-1}{<p,q>+
Q_{\beta}(Ap,p)}\big(\beta^{2}(1/2
+Q_{\beta}(p,q))+\frac{\xi}{2\alpha(\beta^{2})}\big),
\end{equation*}
at the points $\mathfrak{p}(\beta^{2})$ and $(\tau\mathfrak{p})(\beta^{2})$, respectively. Equation (\ref{eq:4.24}) provides the constraint on the discrete potential $a$, denoting it $a=f_{\beta}(p,q)$. Thus, we obtain the following nonlinear map from the linear map (\ref{eq:4.19}):
\begin{equation}\label{eq:4.25}
S_{\beta}: \begin{pmatrix} \tilde{p}\\ \tilde{q}\end{pmatrix}=(A^{2}-\beta^{2})^{-1/2}
\begin{pmatrix}a Ap+\displaystyle\frac{\beta^{2}q}{a-a^{-1}}\\(a-a^{-1})p+a^{-1}A q\end{pmatrix}\Bigg|_{a=f_{\beta}(p,q)}.
\end{equation}
We assert that the map $S_{\beta}$ is an integrable symplectic map sharing the same set of integrals $\{F_{j}(p, q), 1\leq j \leq N\}$ as the Hamiltonian system (\ref{eq:4.12}). In fact, under the constraint (\ref{eq:4.24}) we have
\begin{equation}\label{eq:4.26}
L(\lambda;\tilde{p},\tilde{q})D^{(\beta)}(\lambda;a)=D^{(\beta)}(\lambda;a)L(\lambda;p,q),
\end{equation}
by equations (\ref{eq:4.20}) and (\ref{eq:4.22}), which implies $S_{\beta}^{*}\circ F_{j}(p,q)=F_{j}(p,q),1\leq j\leq N$.

\noindent The symplectic property for $S_{\beta}$, i.e., $S_{\beta}^{*}(\mathrm{d}p\wedge \mathrm{d}q)=\mathrm{d}p\wedge \mathrm{d}q$ relies on the following formula:
\begin{eqnarray}\label{eq:4.27}
\sum\limits_{j=1}^{N} (\mathrm{d}\tilde{p}_{j} \wedge
\mathrm{d}\tilde{q}_{j}-\mathrm{d}p_{j} \wedge \mathrm{d}
q_{j})=\displaystyle\frac{1}{a(a^{2}-1)^{2}}
\mathrm{d}P^{(\beta)}(a; p,q)\wedge \mathrm{d}a.
\end{eqnarray}
As a consequence, a discrete $S^{m}_{\beta}$-flow can be set up by setting $\big(p(m),q(m)\big) =S^{m}_{\beta}(p_{0},q_{0})$,
with $(p_{0},q_{0})$ as an initial point. By equations (\ref{eq:4.6}) and (\ref{eq:4.24}), we denote the corresponding potentials as
\begin{align}\label{eq:4.28}
a(m)=a_{m}=z_{m+1}/z_{m}, \ \ u(m)=u_{m}=z_{m}^{2},
\end{align}
which lead to the discrete spectral problem
\begin{equation}\label{eq:4.29}
h_{\beta}(m+1,\lambda) =
D^{(\beta)}_{m}(\lambda)h_{\beta}(m,\lambda),
\end{equation}
where $D^{(\beta)}_{m}(\lambda) = D^{(\beta)}(\lambda;a_{m}),$ whose fundamental solution matrix $ M_{\beta}(m,\lambda)$ satisfies
\begin{equation}\label{eq:4.30}
 M_{\beta}(m+1,\lambda) =
D^{(\beta)}_{m}(\lambda) M_{\beta}(m,\lambda), \ \
M_{\beta}(0,\lambda) =  I.
\end{equation}
Hence, following this by iteration, we get the solution as a matrix product chain
\begin{equation}\label{eq:4.31}
M_{\beta}(m,\lambda) =
D^{(\beta)}_{m-1}(\lambda)D^{(\beta)}_{m-2}(\lambda)\ldots
D^{(\beta)}_{0}(\lambda),
\end{equation}
which implies $\mathrm{det} M_{\beta}(m,\lambda) = (\lambda^{2}-\beta^{2})^{m}$, and as $\lambda \rightarrow \infty$,
\begin{equation}\label{eq:4.32}
M_{\beta}(m,\lambda)=\begin{pmatrix}\displaystyle\frac{z_{m}}{z_{0}}\lambda^{m}+O(\lambda^{m-2}) & O(\lambda^{m-1})\\
 O(\lambda^{m-1}) & \displaystyle\frac{z_{0}}{z_{m}}\lambda^{m}+O(\lambda^{m-2})\end{pmatrix}.
\end{equation}
Furthermore, from the compatibility relation (\ref{eq:4.26}) along the $S^{m}_{\beta}$-flow
\begin{equation}\label{eq:4.33}
L_{m+1}(\lambda)D^{(\beta)}_{m}(\lambda) =
D^{(\beta)}_{m}(\lambda)L_{m}(\lambda),
\end{equation}
where $L_{m}(\lambda)=L(\lambda;p(m),q(m))$, and equation (\ref{eq:4.31}), we obtain
\begin{align}\label{eq:4.34}
L_{m}(\lambda)M_{\beta}(m,\lambda) =
M_{\beta}(m,\lambda)L_{0}(\lambda),
\end{align}
which is helpful to derive the relevant formulas below for zeros and poles of the corresponding meromorphic functions.

In order to proceed we need some properties of the linear operator $L_{m}(\lambda)$ with values in the solution space of equation, \eqref{eq:4.29}.
Through direct calculation, we obtain the eigenvalues of the operator as follows:
\begin{align}\label{eq:4.35}
&\rho^{\pm}_{\lambda} = \pm\rho_{_{\lambda}} =
\pm\displaystyle\sqrt{-\mathcal{F}_{\lambda}} =
\pm\sqrt{R(\zeta)}/2\lambda\alpha(\zeta),\\ \label{eq:4.36}
&\rho_{\lambda} =\frac{\lambda}{2} (1 + O(\zeta^{-2})), \ \ (\lambda \rightarrow
\infty),
\end{align}
together with the associated eigenfunctions satisfying
\begin{align}\label{eq:4.37}
&h_{\beta,\pm}(m+1,\lambda) =
D^{(\beta)}_{m}(\lambda)h_{\beta,\pm}(m,\lambda),\\ \label{eq:4.38}
&h_{\beta,\pm}(m,\lambda) =
\begin{pmatrix}h_{\beta,\pm}^{(1)}(m,\lambda)\\h_{\beta,\pm}^{(2)}(m,\lambda)\end{pmatrix}
 = M_{\beta}(m,\lambda)
 \begin{pmatrix}c_{\lambda}^{\pm}\\1 \end{pmatrix},\\\label{eq:4.39}
&\big(L_{m}(\lambda)-\rho^{\pm}_{\lambda}\big)h_{\beta,\pm}(m,\lambda)
 = 0.
\end{align}
Let $m=0$ in equations \eqref{eq:4.38} and \eqref{eq:4.39}, then
\begin{equation}\label{eq:4.40}
c^{\pm}_{\lambda}=\displaystyle\frac{L^{11}_{0}(\lambda) \pm
\rho_{_{\lambda}}}{L^{21}_{0}(\lambda)} =
-\displaystyle\frac{L^{12}_{0}(\lambda)}{L^{11}_{0}(\lambda)\mp\rho_{_{\lambda}}},
 \ \ c^{+}_{\lambda}c^{-}_{\lambda} =-
\displaystyle\frac{L^{12}_{0}(\lambda)}{L^{21}_{0}(\lambda)},
\end{equation}
As $\lambda \rightarrow \infty$, we have
\begin{align}\label{eq:4.41}
\begin{split}
&c^{+}_{\lambda}=\lambda(1+ O(\zeta^{-1})),\\
&c^{-}_{\lambda}=<p_{0},q_{0}>\lambda^{-1}(1+ O(\zeta^{-1})).
\end{split}
\end{align}
Moreover, $\lambda c^{+}_{\lambda}$ and $\lambda c^{-}_{\lambda}$ are the values of a meromorphic function on $\mathcal{R}$ constructed by (\ref{eq:4.14}),
\begin{equation*}
\mathcal{C}(\mathfrak{p})=\displaystyle\frac{\zeta/2+\zeta<(\zeta-A^{2})^{-1}p_{0},q_{0}>+\xi/2\alpha(\zeta)}{1+<(\zeta-A^{2})^{-1}Aq_{0},q_{0}>},
\end{equation*}
at the points $\mathfrak{p}(\lambda^{2})$ and $(\tau\mathfrak{p})(\lambda^{2})$, respectively.

Based on the results above, we now prepare some formulas to discuss the common eigenvectors $h_{\beta,\pm}(m,\lambda)$ on the level of Riemann surface theory. Through some calculations, we have
\begin{align}\label{eq:4.42}
\begin{split}
h^{(1)}_{\beta,+}(m,\lambda) \cdot h^{(1)}_{\beta,-}(m,\lambda)
=&<p,q>\mid_{m}
(\zeta-\beta^{2})^{m}\prod\limits_{j=1}^{N}
\displaystyle\frac{\zeta-\mu_{j}^{2}(m)}{\zeta-\nu_{j}^{2}(0)},\\
h^{(2)}_{\beta,+}(m,\lambda) \cdot h^{(2)}_{\beta,-}(m,\lambda)
=&(\zeta-\beta^{2})^{m}\prod\limits_{j=1}^{N}
\displaystyle\frac{\zeta-\nu_{j}^{2}(m)}{\zeta-\nu_{j}^{2}(0)},
\end{split}
\end{align}
by using equations (\ref{eq:4.15}), (\ref{eq:4.31}), (\ref{eq:4.34}), (\ref{eq:4.38}) and (\ref{eq:4.40}). Similarly as the previous sections, the following asymptotic behaviours ($\lambda \rightarrow \infty$) are also obtained via equations (\ref{eq:4.32}), (\ref{eq:4.38}), (\ref{eq:4.41}):
\begin{align}\label{eq:4.43}
\begin{split}
&h_{\beta,+}^{(1)}(m,\lambda)=\displaystyle\frac{z_{m}}{z_{0}}\lambda^{m+1}+O(\lambda^{m-1}),\\
& h_{\beta,-}^{(1)}(m,\lambda)=O(\lambda^{m-1}),\\
&h^{(2)}_{\beta,+}(m,\lambda)=O(\lambda^{m}),\\
&h^{(2)}_{\beta,-}(m,\lambda)=\displaystyle\frac{z_{0}}{z_{m}}\lambda^{m}+O(\lambda^{m-2}).
\end{split}
\end{align}
 Technically, separating out the two cases: $m=2k-1,2k$, from equation (\ref{eq:4.38}) we get
\begin{align}\label{eq:4.44}
\begin{split}
&h_{\beta,\pm}^{(1)}(2k-1,\lambda)=\lambda c^{\pm}_{\lambda}[\lambda^{-1}M_{\beta}^{11}(2k-1,\lambda)]+ M_{\beta}^{12}(2k-1,\lambda),\\
& \lambda h_{\beta,\pm}^{(2)}(2k-1,\lambda)=\lambda c^{\pm}_{\lambda}M_{\beta}^{21}(2k-1,\lambda)+\lambda M_{\beta}^{22}(2k-1,\lambda),\\
&\lambda h_{\beta,\pm}^{(1)}(2k,\lambda)=\lambda c^{\pm}_{\lambda}M_{\beta}^{11}(2k,\lambda)+\lambda M_{\beta}^{12}(2k,\lambda),\\
& h_{\beta,\pm}^{(2)}(2k,\lambda)=\lambda c^{\pm}_{\lambda}[\lambda^{-1}M_{\beta}^{21}(2k,\lambda)]+ M_{\beta}^{22}(2k,\lambda),
\end{split}
\end{align}
then four meromorphic functions on $\mathcal{R}$ can be constructed, with the values at $\mathfrak{p}$ and $\tau\mathfrak{p}$ as
\begin{align}\label{eq:4.45}
\begin{split}
&\mathfrak{h}_{\beta}^{(1)}(2k-1,\mathfrak{p}(\lambda^{2}))=h_{\beta,+}^{(1)}(2k-1,\lambda), \ \
\mathfrak{h}_{\beta}^{(1)}(2k-1,\tau\mathfrak{p}(\lambda^{2}))=h_{\beta,-}^{(1)}(2k-1,\lambda), \\
&\mathfrak{h}_{\beta}^{(2)}(2k-1,\mathfrak{p}(\lambda^{2}))=\lambda h_{\beta,+}^{(2)}(2k-1,\lambda), \ \
\mathfrak{h}_{\beta}^{(2)}(2k-1,\tau\mathfrak{p}(\lambda^{2}))=\lambda h_{\beta,-}^{(2)}(2k-1,\lambda),\\
&\mathfrak{h}_{\beta}^{(1)}(2k,\mathfrak{p}(\lambda^{2}))=\lambda h_{\beta,+}^{(1)}(2k,\lambda), \ \
\mathfrak{h}_{\beta}^{(1)}(2k,\tau\mathfrak{p}(\lambda^{2}))=\lambda h_{\beta,-}^{(1)}(2k,\lambda), \\
&\mathfrak{h}_{\beta}^{(2)}(2k,\mathfrak{p}(\lambda^{2}))= h_{\beta,+}^{(2)}(2k,\lambda), \ \
\mathfrak{h}_{\beta}^{(2)}(2k,\tau\mathfrak{p}(\lambda^{2}))= h_{\beta,-}^{(2)}(2k,\lambda).
\end{split}
\end{align}
From the formulas (\ref{eq:4.42}), we have
\begin{align}\label{eq:4.46}
\begin{split}
&\mathfrak{h}_{\beta}^{(1)}(2k-1,\mathfrak{p}(\lambda^{2}))\mathfrak{h}_{\beta}^{(1)}(2k-1,\tau\mathfrak{p}(\lambda^{2}))
=<p,q>\mid_{2k-1}(\zeta-\beta^{2})^{2k-1}\prod\limits_{j=1}^{N}
\displaystyle\frac{\zeta-\mu_{j}^{2}(2k-1)}{\zeta-\nu_{j}^{2}(0)},\\
&\mathfrak{h}_{\beta}^{(2)}(2k-1,\mathfrak{p}(\lambda^{2}))\mathfrak{h}_{\beta}^{(2)}(2k-1,\tau\mathfrak{p}(\lambda^{2}))
=\zeta(\zeta-\beta^{2})^{2k-1}\prod\limits_{j=1}^{N}
\displaystyle\frac{\zeta-\nu_{j}^{2}(2k-1)}{\zeta-\nu_{j}^{2}(0)},\\
&\mathfrak{h}_{\beta}^{(1)}(2k,\mathfrak{p}(\lambda^{2}))\mathfrak{h}_{\beta}^{(1)}(2k,\tau\mathfrak{p}(\lambda^{2}))
=<p,q>\mid_{2k}\zeta(\zeta-\beta^{2})^{2k}\prod\limits_{j=1}^{N}
\displaystyle\frac{\zeta-\mu_{j}^{2}(2k)}{\zeta-\nu_{j}^{2}(0)},\\
&\mathfrak{h}_{\beta}^{(2)}(2k,\mathfrak{p}(\lambda^{2}))\mathfrak{h}_{\beta}^{(2)}(2k,\tau\mathfrak{p}(\lambda^{2}))
=(\zeta-\beta^{2})^{2k}\prod\limits_{j=1}^{N}
\displaystyle\frac{\zeta-\nu_{j}^{2}(2k)}{\zeta-\nu_{j}^{2}(0)},
\end{split}
\end{align}
Now the zeros and poles for the meromorphic functions $\mathfrak{h}_{\beta}^{(l)}(m,\mathfrak{p})), (l = 1, 2)$ can be obtained by equations (\ref{eq:4.43}) and (\ref{eq:4.46}). This results into the following expressions of the divisors for $\mathfrak{h}_{\beta}^{(l)}(m,\mathfrak{p})), (l = 1, 2)$:
\begin{align}\label{eq:4.47}
\begin{split}
&\mathrm{Div}(\mathfrak{h}_{\beta}^{(1)}(2k-1,\mathfrak{p}))=\sum_{j=1}^{g}\big(\mathfrak{p}(\mu_{j}^{2}(2k-1))-\mathfrak{p}(\nu_{j}^{2}(0))\big)
+ (2k-1)\mathfrak{p}(\beta^{2})-k\infty_{+}-(k-1)\infty_{-},\\
&\mathrm{Div}(\mathfrak{h}_{\beta}^{(2)}(2k-1,\mathfrak{p}))=\sum_{j=1}^{g}\big(\mathfrak{p}(\nu_{j}^{2}(2k-1))-\mathfrak{p}(\nu_{j}^{2}(0))\big)
+ \{\mathfrak{0}\}+(2k-1)\mathfrak{p}(\beta^{2})-k\infty_{+}-k\infty_{-},\\
&\mathrm{Div}(\mathfrak{h}_{\beta}^{(1)}(2k,\mathfrak{p}))=\sum_{j=1}^{g}\big(\mathfrak{p}(\mu_{j}^{2}(2k))-\mathfrak{p}(\nu_{j}^{2}(0))\big)
+\{\mathfrak{0}\}+2k\mathfrak{p}(\beta^{2})-(k+1)\infty_{+}-k\infty_{-},\\
&\mathrm{Div}(\mathfrak{h}_{\beta}^{(2)}(2k,\mathfrak{p}))=\sum_{j=1}^{g}\big(\mathfrak{p}(\nu_{j}^{2}(2k))-\mathfrak{p}(\nu_{j}^{2}(0))\big)
+2k\mathfrak{p}(\beta^{2})-k\infty_{+}-k\infty_{-}.
\end{split}
\end{align}
We now view the above formula (\ref{eq:4.47}) from the Abel-Jacobi variables.  Then, the associated $S^{m}_{\beta}$-flow is linearized on the Jacobi variety $J(\mathcal {R})$ as
\begin{align}\label{eq:4.48}
\begin{split}
&\vec{\psi}(2k-1) \equiv \vec{\phi}(0) +(2k-1)\vec{\Omega}_{\beta}^{-}+k\vec{\Omega},\quad (\mathrm{mod}\mathscr T),\\
&\vec{\phi}(2k-1) \equiv \vec{\phi}(0) +(2k-1)\vec{\Omega}_{\beta}^{-}+k\vec{\Omega}+\vec{\Omega}_{0}^{-},\quad (\mathrm{mod}\mathscr T),\\
&\vec{\psi}(2k) \equiv \vec{\phi}(0) +2k\vec{\Omega}_{\beta}^{-}+(k+1)\vec{\Omega}+\vec{\Omega}_{0}^{-},\quad (\mathrm{mod}\mathscr T),\\
&\vec{\phi}(2k) \equiv \vec{\phi}(0) +2k\vec{\Omega}_{\beta}^{-}+k\vec{\Omega},\quad (\mathrm{mod}\mathscr T),
\end{split}
\end{align}
where $\vec{\Omega}_{\beta}^{-}=\int_{\mathfrak{p}(\beta^{2})}^{\infty_{-}}\vec{\omega}, \vec{\Omega}_{0}^{-}=\int_{\mathfrak{0}}^{\infty_{-}}\vec{\omega},$ and $\vec{\Omega}=\int_{\infty_{-}}^{\infty_{+}}\vec{\omega}$. We can now write down Baker-Akhiezer functions $\mathfrak{h}_{\beta}^{(l)}(m,\mathfrak{p})), (l = 1, 2)$ in terms of theta functions,
\begin{align}\label{eq:4.49}
\begin{split}
\mathfrak{h}_{\beta}^{(1)}(2k-1,\mathfrak{p})=&\displaystyle\frac{
\theta[-\mathscr{A}(\mathfrak{p}) + \vec{\psi}(2k-1) + \vec{K}]}{\theta[-\mathscr{A}(\mathfrak{p}) + \vec{\phi}(0) + \vec{K}]} \cdot
\displaystyle\frac{\theta[-\mathscr{A}(\infty_{+}) + \vec{\phi}(0) + \vec{K}]}{ \theta[-\mathscr{A}(\infty_{+}) + \vec{\psi}(2k-1) + \vec{K}]}\cdot \\ &\cdot\displaystyle\frac{z_{2k-1}}{z_{0}}\cdot\displaystyle\frac{1}{(r_{\beta}^{+})^{k}}\cdot e^{(1-k)\int_{\mathfrak{p}}^{\infty_{+}}\omega[\mathfrak{p}(\beta^{2}),\infty_{-}]+
k\int_{\mathfrak{p}_{0}}^{\mathfrak{p}}\omega[\mathfrak{p}(\beta^{2}),\infty_{+}]}, \\
\mathfrak{h}_{\beta}^{(2)}(2k-1,\mathfrak{p})=&\displaystyle\frac{
\theta[-\mathscr{A}(\mathfrak{p}) + \vec{\phi}(2k-1) + \vec{K}]}{\theta[-\mathscr{A}(\mathfrak{p}) + \vec{\phi}(0) + \vec{K}]}\cdot\displaystyle\frac{
\theta[-\mathscr{A}(\infty_{-}) + \vec{\phi}(0) + \vec{K}]}{\theta[-\mathscr{A}(\infty_{-}) + \vec{\phi}(2k-1) + \vec{K}]}\cdot \\
&\cdot\displaystyle\frac{z_{0}}{z_{2k-1}}\cdot\displaystyle\frac{1}{(r_{\beta}^{-})^{k-1}r_{0}^{-}}\cdot e^{-k\int_{\mathfrak{p}}^{\infty_{-}}\omega[\mathfrak{p}(\beta^{2}),\infty_{+}]+
\int_{\mathfrak{p}_{0}}^{\mathfrak{p}}(k-1)\omega[\mathfrak{p}(\beta^{2}),\infty_{-}]+\omega[\mathfrak{0},\infty_{-}]},\\
\mathfrak{h}_{\beta}^{(1)}(2k,\mathfrak{p})=&\displaystyle\frac{
\theta[-\mathscr{A}(\mathfrak{p}) + \vec{\psi}(2k) + \vec{K}]}{\theta[-\mathscr{A}(\mathfrak{p}) + \vec{\phi}(0) + \vec{K}]} \cdot
\displaystyle\frac{\theta[-\mathscr{A}(\infty_{+}) + \vec{\phi}(0) + \vec{K}]}{ \theta[-\mathscr{A}(\infty_{+}) + \vec{\psi}(2k) + \vec{K}]}\cdot \\ &\cdot\displaystyle\frac{z_{2k}}{z_{0}}\cdot\displaystyle\frac{1}{(r_{\beta}^{+})^{k}r_{0}^{+}}\cdot e^{-k\int_{\mathfrak{p}}^{\infty_{+}}\omega[\mathfrak{p}(\beta^{2}),\infty_{-}]+
\int_{\mathfrak{p}_{0}}^{\mathfrak{p}}k\omega[\mathfrak{p}(\beta^{2}),\infty_{+}]+\omega[\mathfrak{0},\infty_{+}]}, \\
\mathfrak{h}_{\beta}^{(2)}(2k,\mathfrak{p})=&\displaystyle\frac{
\theta[-\mathscr{A}(\mathfrak{p}) + \vec{\phi}(2k) + \vec{K}]}{\theta[-\mathscr{A}(\mathfrak{p}) + \vec{\phi}(0) + \vec{K}]} \cdot
\displaystyle\frac{\theta[-\mathscr{A}(\infty_{-}) + \vec{\phi}(0) + \vec{K}]}{ \theta[-\mathscr{A}(\infty_{-}) + \vec{\phi}(2k) + \vec{K}]}\cdot \\ &\cdot\displaystyle\frac{z_{0}}{z_{2k}}\cdot\displaystyle\frac{1}{(r_{\beta}^{-})^{k}}\cdot e^{-k\int_{\mathfrak{p}}^{\infty_{-}}\omega[\mathfrak{p}(\beta^{2}),\infty_{+}]+
k\int_{\mathfrak{p}_{0}}^{\mathfrak{p}}\omega[\mathfrak{p}(\beta^{2}),\infty_{-}]},
\end{split}
\end{align}
where
\begin{align}\label{eq:4.50}
\begin{split}
&r_{0}^{+}=\underset{\mathfrak{p} \rightarrow \infty^{+}}
\lim\displaystyle\frac{1}{\zeta(\mathfrak{p})}
e^{\int_{\mathfrak{p}_{0}}^{\mathfrak{p}}\omega[\mathfrak{0},\infty_{+}]},\
\ r_{0}^{-}=\underset{\mathfrak{p} \rightarrow \infty^{-}}
{\mathrm{lim}}\displaystyle\frac{1}{\zeta(\mathfrak{p})}
e^{\int_{\mathfrak{p}_{0}}^{\mathfrak{p}}\omega[\mathfrak{0},\infty_{-}]},\\
& r_{\beta}^{+}=\underset{\mathfrak{p} \rightarrow \infty^{+}}
{\mathrm{lim}}\displaystyle\frac{1}{\zeta(\mathfrak{p})}
e^{\int_{\mathfrak{p}_{0}}^{\mathfrak{p}}\omega[\mathfrak{p}(\beta^{2}),\infty_{+}]},\
\  r_{\beta}^{-}= \underset{\mathfrak{p} \rightarrow \infty^{-}}
{\mathrm{lim}}\displaystyle\frac{1}{\zeta(\mathfrak{p})}
e^{\int_{\mathfrak{p}_{0}}^{\mathfrak{p}}\omega[\mathfrak{p}(\beta^{2}),\infty_{-}]}.
\end{split}
\end{align}

With the help of the expression (\ref{eq:4.49}), we now calculate the discrete potential which leads to the finite genus solutions for lSKdV equation (\ref{eq:1.7}).

\noindent{\textbf{Proposition 4.1.}} The discrete potential
$u(m)$, defined by (\ref{eq:4.28}), satisfies the recursive relation,
\begin{align}\label{eq:4.51}
\begin{split}
u(m)-u(m+1)=&u(0)\cdot \displaystyle\frac{
\theta[\vec{\Omega} + \vec{K}(0)]}{\theta[\vec{K}(0)]}
\cdot\displaystyle\frac{\theta[\delta_{m}\vec{\Omega}+(\delta_{m}-\delta_{m+1})\vec{\Omega}_{0}^{-}+\vec{K}(m+1)]}
{\theta[(\delta_{m}+\delta_{m+1})\vec{\Omega}+\vec{K}(m+1)]}\cdot \\
&\cdot \displaystyle\frac{
\theta[\delta_{m+1}\vec{\Omega}-(\delta_{m}-\delta_{m+1})\vec{\Omega}_{0}^{-}+\vec{K}(m)]}{\theta[\vec{\Omega} + \vec{K}(m)]}\cdot(\beta^{2})^{\delta_{m+1}}
\cdot\displaystyle\frac{(r_{\beta}^{+})^{(m+\delta_{m})/2}}{(r_{\beta}^{-})^{(m-\delta_{m})/2}}\cdot\displaystyle\frac{(r^{+}_{0})^{\delta_{m+1}}}
{(r^{-}_{0})^{\delta_{m}}}\cdot\\
&\cdot e^{\int_{\mathfrak{p}_{0}}^{\infty_{+}}\frac{m-\delta_{m}}{2}\omega[\mathfrak{p}(\beta^{2}),\infty_{-}]-
\int_{\mathfrak{p}_{0}}^{\infty_{-}}\frac{m+\delta_{m}}{2}\omega[\mathfrak{p}(\beta^{2}),\infty_{+}]
-\delta_{m+1}\int_{\mathfrak{p}_{0}}^{\mathfrak{p}}\omega[\mathfrak{0},\infty_{+}]
+\delta_{m}\int_{\mathfrak{p}_{0}}^{\mathfrak{p}}\omega[\mathfrak{0},\infty_{-}]},
\end{split}
\end{align}
where $\vec{K}(m)=\vec{\phi}(m)+\vec{K}+\int_{\infty_{+}}^{\mathfrak{p}_{0}}\vec{\omega}$, and $\delta_{j}$ is equal to 0 and 1 for even and odd $j$ respectively.

\noindent\emph{Proof.} From equation (\ref{eq:4.37}), we obtain
\begin{align}\label{eq:4.52}
&\mathfrak{h}_{\beta}^{(1)}(2k+1,\mathfrak{p})=
a_{2k}\mathfrak{h}_{\beta}^{(1)}(2k,\mathfrak{p})+\displaystyle\frac{\beta^{2}}{a_{2k}-a_{2k}^{-1}}
\mathfrak{h}_{\beta}^{(2)}(2k,\mathfrak{p}),\\ \label{eq:4.53}
&\mathfrak{h}_{\beta}^{(1)}(2k+2,\mathfrak{p})=
\zeta a_{2k+1}\mathfrak{h}_{\beta}^{(1)}(2k+1,\mathfrak{p})+\displaystyle\frac{\beta^{2}}{a_{2k+1}-a_{2k+1}^{-1}}
\mathfrak{h}_{\beta}^{(2)}(2k+1,\mathfrak{p}),
\end{align}
which implies
\begin{align}\label{eq:4.54}
&\displaystyle\frac{\mathfrak{h}_{\beta}^{(1)}(2k+1,\mathfrak{p})}{\mathfrak{h}_{\beta}^{(1)}(2k,\mathfrak{p})}=
a_{2k}+\displaystyle\frac{\beta^{2}}{a_{2k}-a_{2k}^{-1}}
\displaystyle\frac{\mathfrak{h}_{\beta}^{(2)}(2k,\mathfrak{p})}{\mathfrak{h}_{\beta}^{(1)}(2k,\mathfrak{p})},\\ \label{eq:4.55}
&\displaystyle\frac{\mathfrak{h}_{\beta}^{(1)}(2k+2,\mathfrak{p})}{\mathfrak{h}_{\beta}^{(1)}(2k+1,\mathfrak{p})}=
\zeta a_{2k+1}+\displaystyle\frac{\beta^{2}}{a_{2k+1}-a_{2k+1}^{-1}}
\displaystyle\frac{\mathfrak{h}_{\beta}^{(2)}(2k+1,\mathfrak{p})}{\mathfrak{h}_{\beta}^{(1)}(2k+1,\mathfrak{p})},
\end{align}
where $\mathfrak{p}=\mathfrak{p}(\zeta), \zeta=\lambda^{2}$. Let $\lambda\rightarrow \beta$, we have
\begin{eqnarray}\label{eq:4.56}
&&a_{2k}^{2}+\beta^{2}
\underset{\lambda \rightarrow \beta}
{\mathrm{lim}}\displaystyle\frac{
\mathfrak{h}_{\beta}^{(2)}(2k,\mathfrak{p}(\lambda^{2}))}{
\mathfrak{h}_{\beta}^{(1)}(2k,\mathfrak{p}(\lambda^{2}))}=1,\\ \label{eq:4.57}
&&a_{2k+1}^{2}+
\underset{\lambda \rightarrow \beta}
{\mathrm{lim}}\displaystyle\frac{
\mathfrak{h}_{\beta}^{(2)}(2k+1,\mathfrak{p}(\lambda^{2}))}{
\mathfrak{h}_{\beta}^{(1)}(2k+1,\mathfrak{p}(\lambda^{2}))}=1,
\end{eqnarray}
according to the divisors given by (\ref{eq:4.47}).

\noindent Then substituting (\ref{eq:4.28}) and the theta function expressions (\ref{eq:4.49}) into (\ref{eq:4.56}) and (\ref{eq:4.57}), respectively, we get
\begin{align}\label{eq:4.58}
\begin{split}
u(2k)-u(2k+1)=&u(0)\cdot \displaystyle\frac{
\theta[-\mathscr{A}(\infty_{-}) + \vec{\phi}(0) + \vec{K}]}{\theta[-\mathscr{A}(\infty_{+}) + \vec{\phi}(0) + \vec{K}]}
\cdot\displaystyle\frac{\theta[-\mathscr{A}(\mathfrak{p}(\beta^{2})) + \vec{\phi}(2k) + \vec{K}]}{\theta[-\mathscr{A}(\mathfrak{p}(\beta^{2})) + \vec{\psi}(2k) + \vec{K}]}\cdot \\
&\cdot \displaystyle\frac{
\theta[-\mathscr{A}(\infty_{+}) + \vec{\psi}(2k) + \vec{K}]}{\theta[-\mathscr{A}(\infty_{-}) + \vec{\phi}(2k) + \vec{K}]}\cdot\beta^{2}\cdot
r_{0}^{+}\cdot\big(\displaystyle\frac{r_{\beta}^{+}}{r_{\beta}^{-}}\big)^{k}\cdot\\
&\cdot e^{k\int_{\mathfrak{p}_{0}}^{\infty_{+}}\omega[\mathfrak{p}(\beta^{2}),\infty_{-}]-
k\int_{\mathfrak{p}_{0}}^{\infty_{-}}\omega[\mathfrak{p}(\beta^{2}),\infty_{+}]-\int_{\mathfrak{p}_{0}}^{\mathfrak{p}}\omega[\mathfrak{0},\infty_{+}]},
\end{split}
\end{align}
and
\begin{align}\label{eq:4.59}
\begin{split}
u(2k+1)-u(2k+2)=&u(0)\cdot \displaystyle\frac{
\theta[-\mathscr{A}(\infty_{-}) + \vec{\phi}(0) + \vec{K}]}{\theta[-\mathscr{A}(\infty_{+}) + \vec{\phi}(0) + \vec{K}]}
\cdot\displaystyle\frac{\theta[-\mathscr{A}(\mathfrak{p}(\beta^{2})) + \vec{\phi}(2k+1) + \vec{K}]}{\theta[-\mathscr{A}(\mathfrak{p}(\beta^{2})) + \vec{\psi}(2k+1) + \vec{K}]}\cdot \\
&\cdot \displaystyle\frac{
\theta[-\mathscr{A}(\infty_{+}) + \vec{\psi}(2k+1) + \vec{K}]}{\theta[-\mathscr{A}(\infty_{-}) + \vec{\phi}(2k+1) + \vec{K}]}\cdot
\displaystyle\frac{1}{r_{0}^{-}}\cdot\displaystyle\frac{(r_{\beta}^{+})^{k+1}}{(r_{\beta}^{-})^{k}}\cdot\\
&\cdot e^{k\int_{\mathfrak{p}_{0}}^{\infty_{+}}\omega[\mathfrak{p}(\beta^{2}),\infty_{-}]-
(k+1)\int_{\mathfrak{p}_{0}}^{\infty_{-}}\omega[\mathfrak{p}(\beta^{2}),\infty_{+}]+\int_{\mathfrak{p}_{0}}^{\mathfrak{p}}\omega[\mathfrak{0},\infty_{-}]},
\end{split}
\end{align}
which give rise to the unified form
\begin{align}\label{eq:4.60}
\begin{split}
u(m)-u(m+1)=&u(0)\cdot \displaystyle\frac{
\theta[-\mathscr{A}(\infty_{-}) + \vec{\phi}(0) + \vec{K}]}{\theta[-\mathscr{A}(\infty_{+}) + \vec{\phi}(0) + \vec{K}]}
\cdot\displaystyle\frac{\theta[-\mathscr{A}(\mathfrak{p}(\beta^{2})) + \vec{\phi}(m) + \vec{K}]}{\theta[-\mathscr{A}(\mathfrak{p}(\beta^{2})) + \vec{\psi}(m) + \vec{K}]}\cdot \\
&\cdot \displaystyle\frac{
\theta[-\mathscr{A}(\infty_{+}) + \vec{\psi}(m) + \vec{K}]}{\theta[-\mathscr{A}(\infty_{-}) + \vec{\phi}(m) + \vec{K}]}\cdot(\beta^{2})^{\delta_{m+1}}
\cdot\displaystyle\frac{(r_{\beta}^{+})^{(m+\delta_{m})/2}}{(r_{\beta}^{-})^{(m-\delta_{m})/2}}\cdot\displaystyle\frac{(r^{+}_{0})^{\delta_{m+1}}}
{(r^{-}_{0})^{\delta_{m}}}\cdot\\
&\cdot e^{\int_{\mathfrak{p}_{0}}^{\infty_{+}}\frac{m-\delta_{m}}{2}\omega[\mathfrak{p}(\beta^{2}),\infty_{-}]-
\int_{\mathfrak{p}_{0}}^{\infty_{-}}\frac{m+\delta_{m}}{2}\omega[\mathfrak{p}(\beta^{2}),\infty_{+}]
-\delta_{m+1}\int_{\mathfrak{p}_{0}}^{\mathfrak{p}}\omega[\mathfrak{0},\infty_{+}]
+\delta_{m}\int_{\mathfrak{p}_{0}}^{\mathfrak{p}}\omega[\mathfrak{0},\infty_{-}]}.
\end{split}
\end{align}
Then by using formulas $-\mathscr{A}(\mathfrak{p}(\beta^{2}))=\vec{\Omega}_{\beta}^{-}+\vec{\Omega}+\int_{\infty_{+}}^{\mathfrak{p}_{0}}\vec{\omega}$ and $\vec{\psi}(m)=\vec{\phi}(m)+\delta_{m+1}\vec{\Omega}+(\delta_{m+1}-\delta_{m})\vec{\Omega}_{0}^{-}$ deduced by equation (\ref{eq:4.48}), equation (\ref{eq:4.51}) is proved.  \hfill $\Box$

\subsection{ The finite genus solutions to the lSKdV equation}

According to the methods used in the preceding sections, we now have two integrable symplectic maps $S_{\beta_{1}}, S_{\beta_{2}}$ by imposing the lattice parameter $\beta$ two values $\beta_{1}, \beta_{2}$ respectively. Then by iteration, $S_{\beta_{1}}^{m}$- and $S_{\beta_{2}}^{n}$-flow commuting with each other are obtained as well. As a result, from recursive relation (\ref{eq:4.51}) we obtain the solutions for the lSKdV equation.

\noindent{\textbf{Proposition 4.2.}} The finite genus solutions for the lSKdV equation (\ref{eq:1.7}) satisfy

\begin{align}\label{eq:4.61}
\begin{split}
u(m,n)-u(m+1,n)=&u(0,n)\cdot \displaystyle\frac{
\theta[\vec{\Omega} + \vec{K}(0,n)]}{\theta[\vec{K}(0,n)]}
\cdot\displaystyle\frac{\theta[\delta_{m}\vec{\Omega}+(\delta_{m}-\delta_{m+1})\vec{\Omega}_{0}^{-}+\vec{K}(m+1,n)]}
{\theta[(\delta_{m}+\delta_{m+1})\vec{\Omega}+\vec{K}(m+1,n)]}\cdot \\
&\cdot \displaystyle\frac{
\theta[\delta_{m+1}\vec{\Omega}-(\delta_{m}-\delta_{m+1})\vec{\Omega}_{0}^{-}+\vec{K}(m,n)]}{\theta[\vec{\Omega} + \vec{K}(m,n)]}\cdot(\beta_{1}^{2})^{\delta_{m+1}}
\cdot \\
&\cdot\displaystyle\frac{(r_{\beta_{1}}^{+})^{(m+\delta_{m})/2}}{(r_{\beta_{1}}^{-})^{(m-\delta_{m})/2}}\cdot\displaystyle\frac{(r^{+}_{0})^{\delta_{m+1}}}
{(r^{-}_{0})^{\delta_{m}}}\cdot\\
&\cdot e^{\int_{\mathfrak{p}_{0}}^{\infty_{+}}\frac{m-\delta_{m}}{2}\omega[\mathfrak{p}(\beta_{1}^{2}),\infty_{-}]-
\int_{\mathfrak{p}_{0}}^{\infty_{-}}\frac{m+\delta_{m}}{2}\omega[\mathfrak{p}(\beta_{_{1}}^{2}),\infty_{+}]
-\delta_{m+1}\int_{\mathfrak{p}_{0}}^{\mathfrak{p}}\omega[\mathfrak{0},\infty_{+}]
+\delta_{m}\int_{\mathfrak{p}_{0}}^{\mathfrak{p}}\omega[\mathfrak{0},\infty_{-}]},
\end{split}
\end{align}
where $u(0,n)$ is given by
\begin{align}\label{eq:4.62}
\begin{split}
u(0,n)-u(0,n+1)=&u(0,0)\cdot \displaystyle\frac{
\theta[\vec{\Omega} + \vec{K}(0,0)]}{\theta[\vec{K}(0,0)]}
\cdot\displaystyle\frac{\theta[\delta_{n}\vec{\Omega}+(\delta_{n}-\delta_{n+1})\vec{\Omega}_{0}^{-}+\vec{K}(0,n+1)]}
{\theta[(\delta_{n}+\delta_{n+1})\vec{\Omega}+\vec{K}(0,n+1)]}\cdot \\
&\cdot \displaystyle\frac{
\theta[\delta_{n+1}\vec{\Omega}-(\delta_{n}-\delta_{n+1})\vec{\Omega}_{0}^{-}+\vec{K}(0,n)]}{\theta[\vec{\Omega} + \vec{K}(0,n)]}\cdot(\beta_{2}^{2})^{\delta_{n+1}}
\cdot\displaystyle\frac{(r_{\beta_{2}}^{+})^{(n+\delta_{n})/2}}{(r_{\beta_{2}}^{-})^{(n-\delta_{n})/2}}\cdot\displaystyle\frac{(r^{+}_{0})^{\delta_{n+1}}}
{(r^{-}_{0})^{\delta_{n}}}\cdot\\
&\cdot e^{\int_{\mathfrak{p}_{0}}^{\infty_{+}}\frac{n-\delta_{n}}{2}\omega[\mathfrak{p}(\beta_{2}^{2}),\infty_{-}]-
\int_{\mathfrak{p}_{0}}^{\infty_{-}}\frac{n+\delta_{n}}{2}\omega[\mathfrak{p}(\beta_{2}^{2}),\infty_{+}]
-\delta_{n+1}\int_{\mathfrak{p}_{0}}^{\mathfrak{p}}\omega[\mathfrak{0},\infty_{+}]
+\delta_{n}\int_{\mathfrak{p}_{0}}^{\mathfrak{p}}\omega[\mathfrak{0},\infty_{-}]},
\end{split}
\end{align}
and $\vec{K}(m,n) = \vec{\phi}(m,n) + \vec{K} +
\int_{\infty_{+}}^{\mathfrak{p}_{0}}\vec{\omega}, \ \ \vec{\phi}(m,n) =\vec{\phi}(0,0)+
m\vec{\Omega}_{\beta_{1}}^{-}+n\vec{\Omega}_{\beta_{2}}^{-}+\displaystyle\frac{m+n+\delta_{m}+\delta_{n}}{2}\vec{\Omega}
+(\delta_{m}+\delta_{n})\vec{\Omega}_{0}^{-}$. Besides, $r_{\beta_{j}}^{+}, r_{\beta_{j}}^{-}$ are obtained by putting $\beta=\beta_{j}, j=1,2$ in equation
(\ref{eq:4.50}) respectively.

Another way to obtain the analytic solution in terms of theta functions for equation (\ref{eq:1.7}) is calculating the potential $u(m)$ by the discrete integration
\begin{equation*}
u(m)=u(0)+\sum_{j=1}^{m}\big(u(j)-u(j-1)\big),
\end{equation*}
with the help of equation (\ref{eq:4.51}). The evolution of $u(m)$ along the corresponding discrete flows leads to the solutions as well.

\section{ Conclusion}

In this paper, we exhibited a new version of the algebro-geometric approach to deal with the partial difference equations of KdV-type, which is different from the existing results in the literatures \cite{Cao,Cao3,Cao5}.

We have presented examples of integrable symplectic maps and finite genus solutions for lattice KdV-type equations. In the lpKdV and lSKdV cases, there are two discrete potentials,
and we need to impose constraints between them in order to construct the algebro-geometric solutions using the technique of nonlinearisation. Applying the method and the
constraints, we end up with expressions for a single potential for the lSKdV equation as in the lpmKdV case. These cases share a similar algebro-geometric  structure when constructing
the explicit solutions in terms of theta functions. However, in the lpKdV case, the Riemann surface is different and the constraint is not enough to characterize the
solution. Hence, an alternative parametrization was constructed in order to solve the problem.

In this paper the discrete version of the Liouville-Arnol'd theorem, \cite{Veselov,Bruschi}, plays an essential role. We point out in the present context that different Liouville integrable reductions can be considered associated with distinct Hamiltonian systems, leading all to solutions of one and the same partial difference equation.

At this juncture, we would like to point out that Noether's principle for Hamiltonian systems tells us that there is a correspondence between integrals and symmetries.
Furthermore, the integrals of a Hamiltonian system form a Lie algebra with respect to the Poisson bracket, while the corresponding flows generate a Lie group. Therefore, we may
conjecture that the algebraic structure behind the approach employed in our analysis could shed a light
on this phenomenon in discrete integrable systems.

\vspace*{0.5cm} \noindent{\bf Acknowledgments}

This work is supported by National Natural Science Foundation of China (Grant
Nos. 11426206; 11501521), State Scholarship Found of China (CSC No. 201907045035), and Graduate Student Education Research Foundation
of Zhengzhou University (Grant No. YJSXWKC201913). We would like to express many thanks to Prof. Cewen Cao and Prof. Da-jun Zhang for helpful discussions.

\end{document}